\documentclass[lettersize,journal]{IEEEtran}
\usepackage{amsmath,amsfonts}
\usepackage{amssymb}
\usepackage{amsthm}

\usepackage{array}

\usepackage{graphicx}
\usepackage{subfig}  
\usepackage{caption} 
\usepackage{stfloats}

\usepackage{enumitem}
\usepackage{textcomp}

\usepackage{cite}
\UseRawInputEncoding

\begin{document}

\title{Parameterized TDOA: 
TDOA Estimation for Mobile Target Localization
in a Time-Division Broadcast Positioning System}
\author{Chenxin~Tu,
        Xiaowei~Cui,
        Gang~Liu,
        Sihao Zhao,~\IEEEmembership{Senior~Member,~IEEE},
        and Mingquan Lu 
\thanks{{Corresponding authors: Xiaowei Cui; Mingquan Lu}.}
\thanks{Chenxin Tu and Gang Liu are with the Department of Electronic Engineering, Tsinghua
University, Beijing 100084, China (e-mail: tcx22@mails.tsinghua.edu.cn; liu\_gang@tsinghua.edu.cn).}
\thanks{Xiaowei Cui and Mingquan Lu are with the Department of Electronic Engineering, State Key Laboratory of Space Network and Communication, Tsinghua University, Beijing 100084, China (e-mail:
cxw2005@tsinghua.edu.cn; lumq@tsinghua.edu.cn).}
\thanks{Sihao Zhao is an independent researcher in Alberta, Canada (e-mail: zsh01@tsinghua.org.cn).}
\thanks{\copyright2025 IEEE. Personal use of this material is permitted. Permission from IEEE must be obtained for all other uses, in any current or future media, including reprinting/republishing this material for advertising or promotional purposes, creating new collective works, for resale or redistribution to servers or lists, or reuse of any copyrighted component of this work in other works.}
}



\maketitle

\begin{abstract}
In a time-division broadcast positioning system (TDBPS), localizing mobile targets using classical time difference of arrival (TDOA) methods poses significant challenges. Concurrent TDOA measurements are infeasible because targets receive signals from different anchors and extract their transmission times at different reception times, as well as at varying positions. Traditional TDOA estimation schemes implicitly assume that the target remains stationary during the measurement period, which is impractical for mobile targets exhibiting high dynamics. Existing methods for mobile target localization are mostly specialized and rely on motion modeling and do not rely on the concurrent TDOA measurements. This issue limits their direct use of the well-established classical TDOA-based localization methods and complicating the entire localization process. In this paper, to obtain concurrent TDOA estimates at any instant out of the sequential measurements for direct use of existing TDOA-based localization methods, we propose a novel TDOA estimation method, termed parameterized TDOA (P-TDOA). By approximating the time-varying TDOA as a polynomial function over a short period, we transform the TDOA estimation problem into a model parameter estimation problem and derive the desired TDOA estimates thereafter. Theoretical analysis shows that, under certain conditions, the proposed P-TDOA method closely approaches the Cram\'er-Rao Lower Bound (CRLB) for TDOA estimation in concurrent measurement scenarios, despite measurements being obtained sequentially. Extensive numerical simulations validate our theoretical analysis and demonstrate the effectiveness of the proposed method, highlighting substantial improvements over existing approaches across various scenarios.
\end{abstract}

\section{Introduction}
\IEEEPARstart{W}{ireless} positioning has found widespread applications across various fields, including industrial automation \cite{martalo2021improved}, healthcare \cite{wyffels2014distributed}, transportation \cite{zhang2019rfid}, and military combat \cite{alhmiedat2012study}. 
Among various types of positioning systems, time difference of arrival (TDOA)-based broadcast systems, also known as downlink TDOA (DL-TDOA), have gained more attention and applications due to their privacy in the positioning process and scalability in terms of user capacity\cite{wen2020automated,you2021vehicles,santoro2023uwb}. In such systems, several anchors with known positions and synchronized clocks continuously broadcast ranging signals containing transmission time information according to a designated multiple access scheme, while one or more asynchronous targets to be localized passively listen to these ranging signals and extract corresponding transmission times at specific reception times. In global navigation satellite systems (GNSS) using code division multiple access (CDMA), such as the well-known global positioning system (GPS) and BeiDou satellite navigation system (BDS)\cite{spilker1978gps,kaplan2017understanding,yang2019introduction}, the target can extract the transmission times from multiple different anchors (i.e. satellites) at the same reception time. Then several concurrent TDOA measurements can be derived by subtracting the transmission time of a selected reference anchor from those of other anchors\footnote{
Unless otherwise specified, TDOA in this paper refers to the difference of signal propagation times between a target and an anchor pair at the same instant,  corresponding to the distance difference if multiplied by the signal propagation speed. Furthermore, concurrent TDOAs refer to a set of such TDOAs between a target and multiple anchor pairs at the same instant.
}, 
which eliminates the impact of the target's clock offset relative to the system. Subsequently, the target's location can be determined by solving a set of hyperbolic equations formed by these TDOA measurements, with various classical methods having been extensively studied across different scenarios\cite{chan1994simple,sun2011asymptotically,ghany2019parametric,kravets2024new}.

However, for ground-based or indoor positioning systems, CDMA suffers from the near-far effect due to significant variations in the relative distances between the target and different anchors\cite{picois2014near}. Consequently, time division multiple access (TDMA), which is immune to the near-far effect, is often adopted in these scenarios, forming the so-called time division broadcast positioning systems (TDBPS) exemplified by ultra-wideband (UWB) device networks\cite{chehri2009uwb,zwirello2012uwb} and acoustic sensor networks\cite{toky2020localization} deployed in indoor or underwater environments. 
In these systems, time division causes anchors to broadcast ranging signals sequentially in their respective time slots, while targets can only extract the transmission times of different anchors at different reception times, which complicates the derivation of TDOA measurements.

In scenarios where the target is stationary or moves slowly, it is assumed that the position of the target at different reception times remains unchanged. 
In order to obtain the TDOA measurements between the target and any pair of anchors, we need to compensate for the clock offset caused by the target's clock drift between the two reception times,
which is equivalent to aligning the reception times of the ranging signals from different anchors to a virtual reception time, ultimately resulting in concurrent measurements. 
The core problem lies in estimating the target's relative clock drift, and there are mainly two approaches to address it.
Dotlic \textit{et al.} measure the carrier frequency offset (CFO) from the received signals to estimate the relative clock drift in uplink TDOA \cite{dotlic2018ranging}, 
and P{\u{a}}tru \textit{et al.} extend this approach to downlink TDOA scenarios \cite{puatru2023flextdoa}. 
Zhang \textit{et al.} \cite{zhang2019high, zhang2021signal} use the ratio between the reception intervals of the target and the transmission intervals of the reference anchor over the same period to estimate the relative clock drift.

In contrast to the stationary or low-motion cases, a more challenging scenario arises when the target moves rapidly, leading to significant changes in its position at different reception times. Time-varying distances between the target and anchors, combined with other adverse factors including different reception times and the target's clock drift, hinder traditional methods from reconstructing ideal concurrent TDOA measurements similar to GNSS based on the extracted transmission times. Most existing research for mobile target localization in a TDBPS has abandoned TDOA estimation, relying directly on the recorded transmission and reception times and modeling the target's motion and clock offset to jointly estimate parameters including position, velocity, clock drift, and clock offset\cite{shi2020sequential,zhao2021optimal,guo2022new}. Shi \textit{et al.} \cite{shi2020sequential} propose an extended two-step weighted least squares (WLS) method which reparameterizes the localization problem into a higher-dimensional space. Zhao \textit{et al.} \cite{zhao2021optimal} develop a maximum likelihood (ML) based method that performs joint localization and synchronization through iterative processes. However, this iterative method requires a good initial guess. Guo \textit{et al.} \cite{guo2022new} improve upon previous work by deriving a closed-form solution, eliminating the need for an initial guess. Nevertheless, the introduction of simultaneous quadratic equations may lead to cases where no real solutions exist, resulting in potential failures. Moreover, these methods all require the introduction of additional parameters to be estimated and intermediate variables, which not only reduces the positioning accuracy but also increases the number of required anchors. For instance, the approach in \cite{shi2020sequential} requires a minimum of nine anchors, while \cite{guo2022new} requires at least seven, in 2D scenarios, making these methods impractical for many real-world applications. Additionally, the iterative or complex nonlinear computations involved result in high computational complexity. Different from aforementioned methods which abandon TDOA estimation, a timestamp conversion-based method is proposed in \cite{sun2024novel}, which employs Lagrange interpolation to convert sequential measurements to a unified reference instant, effectively transforming the problem into a GNSS-like scenario with concurrent measurements. However, the Lagrange interpolation is equivalent to linear interpolation without higher-order measurements, which implicitly assumes constant velocity for the mobile target during the measurement period, imposing restrictions on protocol designs and potentially leading to performance degradation in highly dynamic scenarios.

In order to tackle the challenge of mobile target localization in a TDBPS, in this paper we propose a new method for TDOA estimation based on sequential measurements, namely, parameterized TDOA (P-TDOA). Inspired by \cite{rajan2015joint} and \cite{sahu2024data}, which parameterize the time-varying distance and space-varying TDOA, respectively, we model the time-varying TDOA between a target and any pair of anchors as a polynomial function of time. This new approach reformulates the TDOA estimation problem into a model parameter estimation problem. Unlike the method in \cite{rajan2015joint}, which relies on two-way ranging, our approach utilizes one-way ranging, making it well-suited for passive targets in a TDBPS. Furthermore, while the space-varying TDOA in \cite{sahu2024data} accounts for variations in signal propagation speed through nonhomogeneous media, it does not address challenges associated with target mobility, which our new method explicitly tackles. After modeling the time-varying TDOA as a polynomial function of time over a short period, we derive the equations based on recorded transmission and reception times in terms of the TDOA model parameters through dedicated transformation. Then a WLS approach termed mobile weighted least squares (MWLS) is developed to solve this estimation problem. In addition, we design a strategy for constructing these equations, called the successive time difference strategy (STDS), to ensure that the equations in terms of the TDOA model parameters are valid and the problem is well-posed. Finally, the concurrent TDOA estimates between the target and different anchor pairs can be obtained based on the estimated TDOA models with the same time assigned.

The main contributions of this paper are summarized as follows:
\begin{itemize}
    \item We propose a novel method for TDOA estimation for mobile targets in a TDBPS called P-TDOA, through which we can obtain concurrent TDOA estimates and directly apply classical TDOA-based localization methods for mobile targets. 
    \item Theoretical analysis shows that, under certain conditions, TDOA estimation by P-TDOA in sequential measurement settings can asymptotically approach the CRLB in concurrent measurement scenarios.
    \item Simulation results validate the theoretical analysis and demonstrate significant superiority of our method over existing methods across various scenarios.
\end{itemize}

The rest of the paper is organized as follows. In Section \textrm{II}, we formulate our problem, including the basic settings of a TDBPS, the time-division (TD) measurement protocol adopted and the measurement models for transmission and reception times as well as the asynchronous clocks of targets. The significant challenges of TDOA estimation for asynchronous mobile targets are presented in Section \textrm{III}. Next, the proposed P-TDOA method is detailed in Section \textrm{IV}. In Section \textrm{V}, we derive the CRLB of TDOA estimation in concurrent measurement scenarios and the theoretical mean square error (MSE) of P-TDOA in our problem settings. Extensive simulations are presented in Section \textrm{VI} to validate the effectiveness of the proposed method for TDOA estimation and demonstrate its advantages for mobile target localization. Finally, our conclusions are summarized in Section \textrm{VII}.

The main notations in the paper are summarized in Table \textrm{I}.

\begin{table}[!t]
\centering
\caption{Notation List}
\begin{tabular}{p{2.1cm} p{6cm}}
\hline
\textbf{Notation} & \textbf{Description} \\
\hline
lowercase $x$ & scalar \\
bold lowercase $\boldsymbol{x}$ & vector \\
bold uppercase $\boldsymbol{X}$ & matrix \\
$\left\lVert \boldsymbol{x} \right\rVert$ & Euclidean norm of a vector \\
$\left[ \boldsymbol{X} \right]_{i,j}$ &  entry at the $i$-th row and the $j$-th column of a matrix\\
$\mathcal{V}$ & an operator that converts a vector into the transpose of a Vandermonde matrix  \\
$\mathbb{E}[\cdot]$ & expectation operator \\
$\operatorname{diag}(\cdot)$ & diagonal matrix with the entries inside \\
$\boldsymbol{V}^{\dagger}$ & left inverse of matrix 
$\boldsymbol{V}$ \\
$(\cdot)^{T}$ & transpose operator \\
$(\cdot)^{\odot N}$ & element-wise matrix exponent \\
$\hat{(\cdot)}$ & The term with a hat symbol above indicates that it is an estimate contaminated with noise. \\
$\succeq$ & Greater than or equal in the L\"{o}wner partial order. For example, $\boldsymbol{A} \succeq \boldsymbol{B}$ indicates that $\boldsymbol{A}-\boldsymbol{B}$ is a positive semidefinite matrix. \\
$\boldsymbol{I}_M$ & $M \times M$ identity matrix \\
$\boldsymbol{0}_M$ & $M \times M$ zero matrix \\
$\mathbb{R}^{M\times1}$ & $M \times 1$  real-valued vector \\

$K$    & dimension of the studied scenario ($K=2 \text{ or } 3$)\\
$i,j$  & indices of anchors \\
$u$    & index of the target \\
$N_a$, $\mathcal{N}_a$  & number of anchors and the set of anchors' indices \\
$\omega_u$, $\phi_u$   & unknown clock drift and clock offset of target $u$ \\ 
$\alpha_u$, $\beta_u$  & unknown clock calibration parameters of target $u$ \\
$T_f$, $T_s$       & the duration of one frame and the duration of one time slot \\
$N_f$, $N_s$       & the number of frames in a solution period and the number of time slots in a single frame \\
$t$, $T$  &  The lowercase $t$ represents the system time while the uppercase $T$ represents the local time. The specific local clock for a particular anchor or target, as well as the specific time instance, are indicated by the subscripts and superscripts. \\
$T_{i}^{(m)}, v_{i}^{(m)} ,\sigma_{t}$
& The transmission time of anchor $i$ in frame $m$, its measurement noise, and the deviation of the noise.  \\
$T_{ui}^{(m)}, w_{ui}^{(m)}, \sigma_{r}$ 
& The reception time of target $u$ for the signal from anchor $i$ in frame $m$, its measurement noise, and the deviation of the noise.  \\
$\tau_{ui}$, $\tau_{ui}$, $\tau_{u;ij}$ &  The propagation times from anchor $i$ and anchor $j$ to target $u$, as well as their difference (TDOA).  \\
$L$, $l$ & The number of parameters in the polynomial model and the index of the parameter (start from 0 to $L-1$). \\
$\boldsymbol{\gamma}_{ui}$, $\boldsymbol{\gamma}_{uj}$, $\boldsymbol{\gamma}_{u;ij}$ &  The vectors of model parameters of the propagation times from anchor $i$ and anchor $j$ to target $u$, and the vector of model parameters of the TDOA between target $u$ and anchor pair $(i,j)$. \\
$a_{l,ui}^{(m,n,p,q)}$, $a_{l,uj}^{(m,n,p,q)}$, $a_{l,u;ij}^{(m,n,p,q)}$ &  The coefficients of the $l$-th parameter in $\boldsymbol{\gamma}_{ui}$ (${\gamma}_{l,ui}$), $\boldsymbol{\gamma}_{uj}$ (${\gamma}_{l,uj}$) and $\boldsymbol{\gamma}_{u;ij}$ (${\gamma}_{l,u;ij}$), respectively. The superscript $(m, n, p, q)$ indicates
association with the $(m, n, p, q)$-th message. \\
\hline
\end{tabular}
\end{table}

\section{Problem Formulation} \label{problemFormulation}

\begin{figure}
    \centering
    \includegraphics[width=0.45\textwidth]{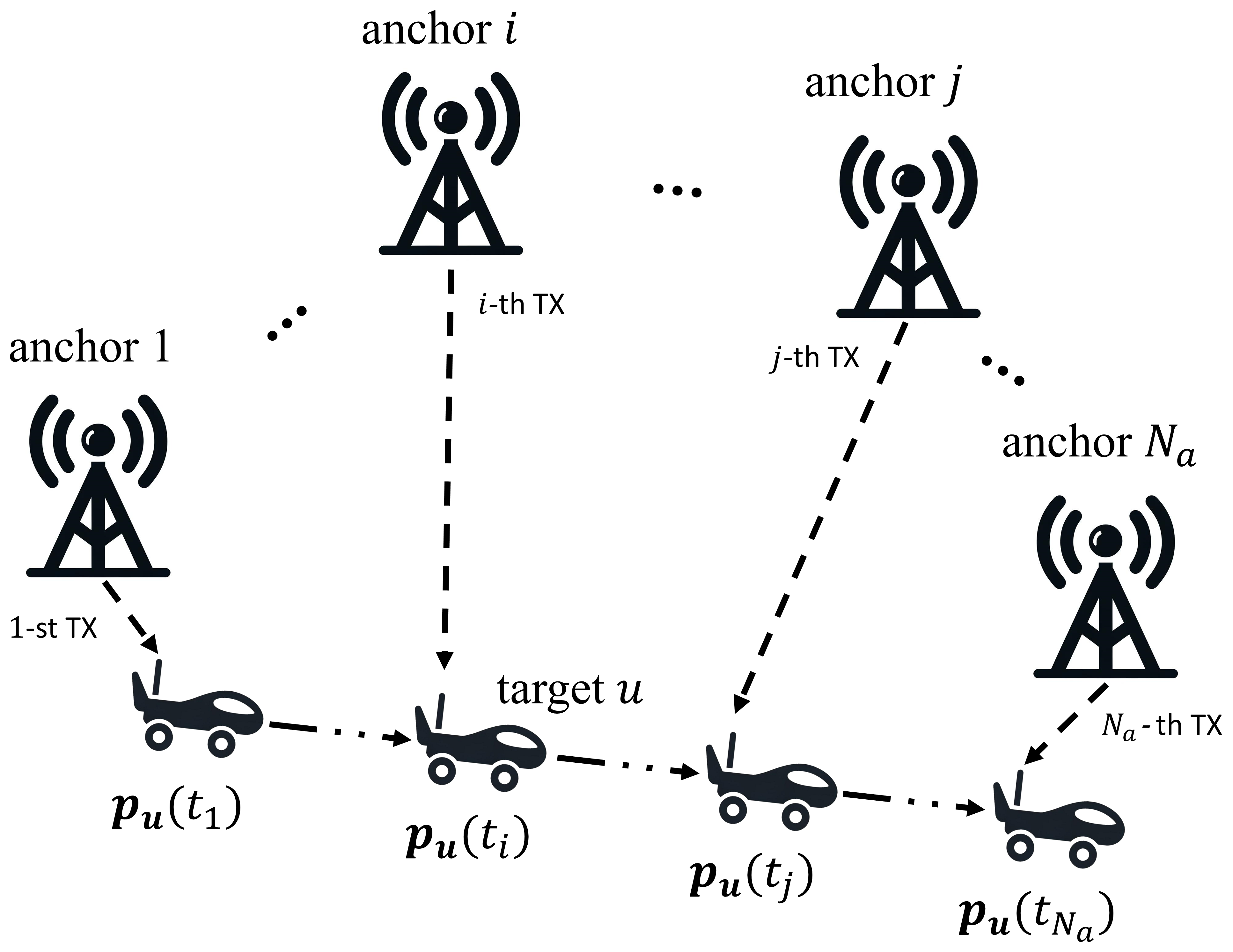}
    \caption{
    Sequential signal transmission and reception in a TDBPS, where anchors broadcast signals sequentially, and the target passively receives them. The position and clock offset of the mobile target change over time.
    }
    \label{fig1}
\end{figure}

As shown in Fig. \ref{fig1}, we consider a TDBPS in a $K$-dimensional ($K=2 \text{ or } 3$) scenario,
consisting of $N_a$ stationary anchors with known positions indexed from 1 to $N_a$. All anchors are assumed to be well synchronized to a common time reference, referred to as the system time and denoted as $t$, through wired or wireless means\cite{yu2009ground,mcelroy2014comparison,leugner2016comparison}. To simplify the illustration, we focus on a single mobile target, denoted by $u$.  

\begin{figure}[t]
    \centering
    \includegraphics[width=0.5\textwidth]{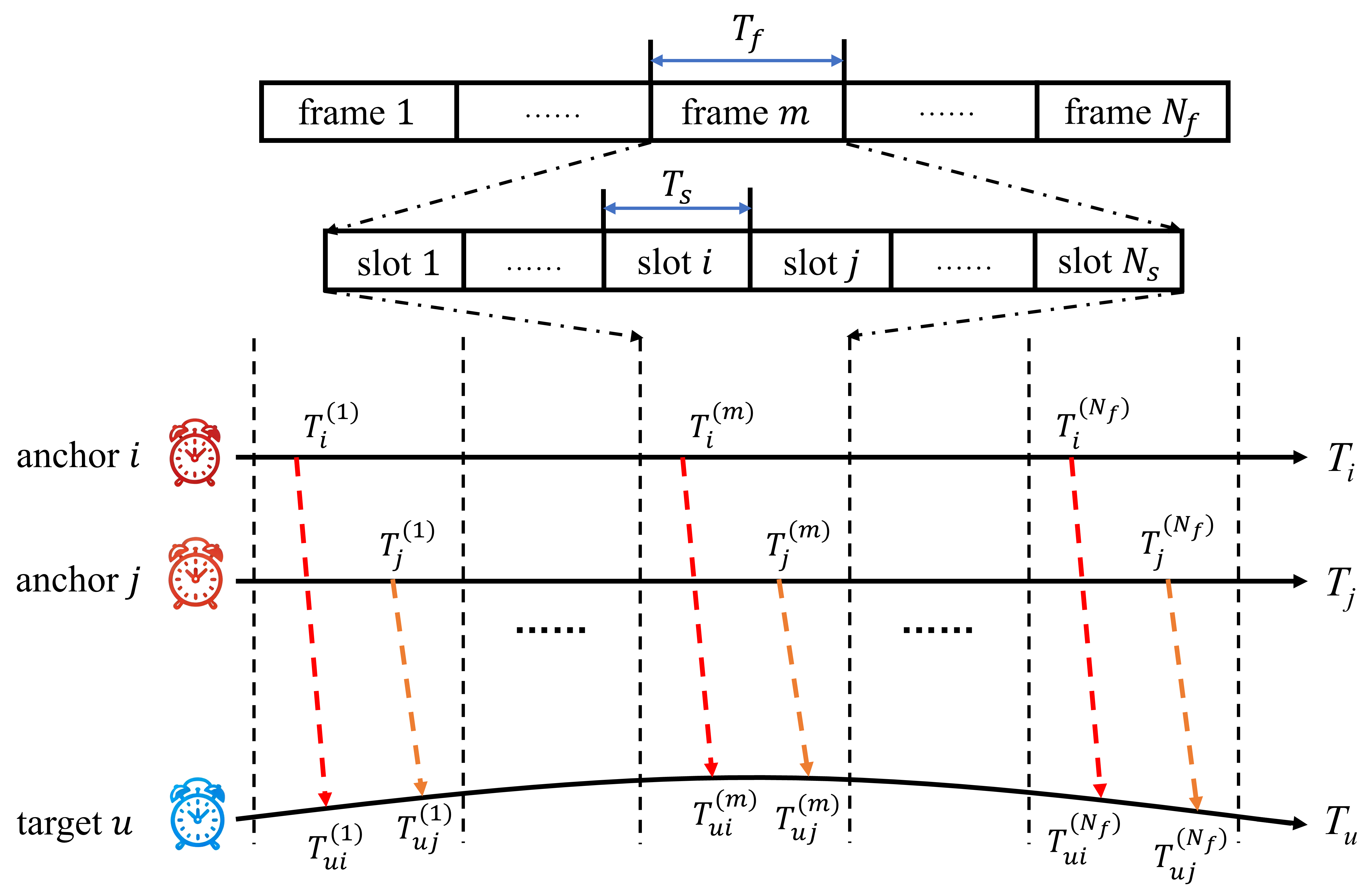}
    \caption{The diagram of periodic broadcast protocol. Each anchor occupies an individual time slot to broadcast signals. The curved lines of target $u$ symbolize the influence of its asynchronous clock and motion.}
    \label{fig2}
\end{figure}

Anchors sequentially broadcast signals according to the protocol shown in Fig. \ref{fig2}\footnote{It is worth mentioning that our proposed P-TDOA method is not restricted to the periodic broadcast protocol but can function under other TD measurement protocols as well. We adopt this protocol for its widespread applications and ease of illustration. }, which is similar to those described in \cite{zhang2019high,zhang2021signal,shi2020blas}, and target $u$ passively receives these signals.
The minimal operational period, referred to as a frame, has a duration denoted as $T_f$. Each frame is subdivided into $N_s$ time slots ($N_s \geq N_a$) whose duration is $T_s$, and each anchor occupies an individual slot to broadcasts signals. 

Through signal processing, the target can estimate the arrival times of signals broadcast by different anchors, referred to as reception times, and extract the corresponding departure times from the anchors, referred to as transmission times. Generally, the noise in transmission and reception time measurements differs in magnitude; thus, we denote them by $v$ and $w$, respectively.
As annotated in Fig. \ref{fig2}, the transmission time of anchor $i$ in frame $m$ and the corresponding reception time recorded by target $u$ are modeled as follows,
\begin{equation}
\begin{aligned}
    \hat{T}_i^{(m)} &= {T}_i^{(m)} + v_{i}^{(m)} \\
    \hat{T}_{ui}^{(m)} &= {T}_{ui}^{(m)} + w_{ui}^{(m)}         
\end{aligned}
\label{eq1}
\end{equation}
where the superscript $(m)$ denotes that these items belong to frame $m$. $\hat{T}_i^{(m)}$, ${T}_i^{(m)}$ and $v_i^{(m)}$ represent the measured transmission time in anchor $i$'s local time, its noise-free actual value and its measurement noise, respectively. Similarly, $\hat{T}_{ui}^{(m)}$, ${T}_{ui}^{(m)}$ and $w_{ui}^{(m)}$ represent the measured reception time in target $u$'s local time, its noise-free actual value and its measurement noise, respectively. Typically, the measurement noise of transmission and reception times is assumed to follow a Gaussian distribution \cite{zhang2019high,zhang2021signal}. 

It's worth mentioning that both transmission and reception times are recorded by the anchors and the target according to their respective clocks.
While the anchors are well synchronized to the system time, the target remains asynchronous, meaning that the differences between reception and transmission times do not directly correspond to the signal propagation times unless the target's local time is aligned with the system time. Although the deviation of the target's local time from the system time is inherently nonlinear, for typical clock oscillators with small Allan deviations which are common in practice, it can be effectively modeled as a linear error over a short period \cite{freris2010fundamental}. In this way, the relationship between target $u$'s local time $T_u$ and system time $t$ is modeled as
\begin{gather}
     T_u = \mathcal{G}(t) = \omega_u t+\phi_u \notag \\
    t  = \mathcal{H}(T_u) = \alpha_u T_u + \beta_u
    \label{eq2}
\end{gather}
where $\omega_u$ and $\phi_u$ denote the clock drift and the initial clock offset of target $u$, respectively, $\left[\alpha_u, \beta_u \right] \triangleq [\omega_u^{-1}, -\phi_u\omega_u^{-1}]$ are the calibration parameters, where we name $\alpha_u$ and $\beta_u$ as the drift calibration parameter and the offset calibration parameter, respectively. In particular, if target $u$ is well synchronized with the system time, we have $[\omega_u, \phi_u]=[1,0]$ and $[\alpha_u, \beta_u]=[1,0]$.

In this paper, we consider $N_f$ frames as a solution period. 
To localize the mobile target in a TDBPS, we collect the transmission times from multiple anchors and corresponding reception times from the target across the solution period.
We aim to estimate the TDOA between target $u$ and any pair of anchors at any moment over the solution period, thereby obtaining concurrent TDOA measurements from multiple anchor pairs.

\section{TDOA Estimation in a TDBPS} \label{challenge}

\begin{figure}
    \centering
    \includegraphics[width=\linewidth]{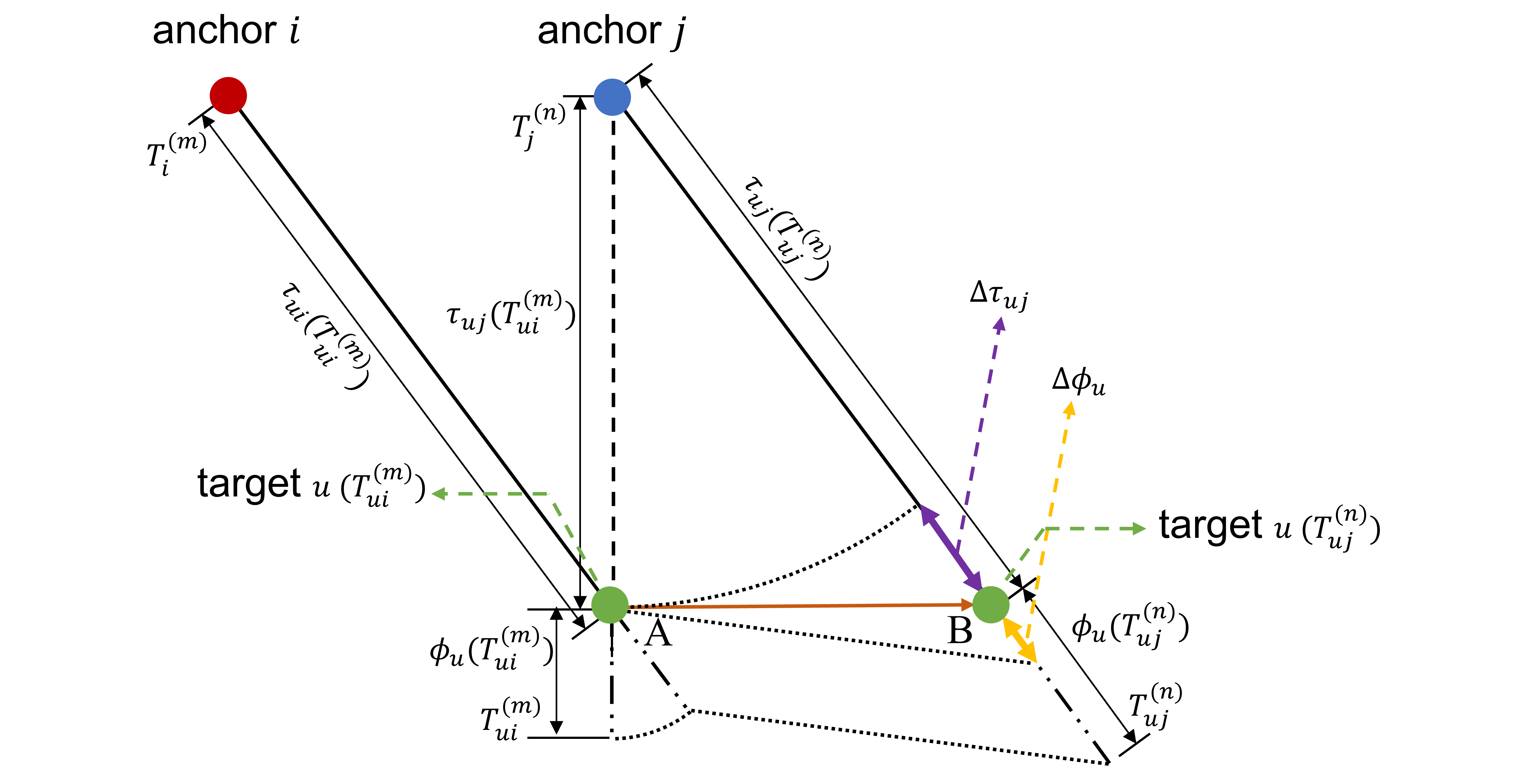}
    \caption{TDOA estimation for an asynchronous mobile target in a TDBPS. The black solid lines represent the actual signal propagation times, while the black dashed lines indicate the virtual signal propagation times. The dash-dot lines represent the influence of clock offsets of target $u$.}
    \label{fig3}
\end{figure}

In this section, we demonstrate the significant challenges of TDOA estimation for an asynchronous mobile target in a TDBPS. As shown in Fig. \ref{fig3}, anchor $i$ and anchor $j$ broadcast signals sequentially at $T_i^{(m)}$ and $T_j^{(n)}$, and target $u$ receives these signals at $T_{ui}^{(m)}$ and $T_{uj}^{(n)}$, respectively. During the reception interval, target $u$ moves from Point $A$ to Point $B$, resulting in the variation of distances between target $u$ and anchor $i$ and anchor $j$. The transmission and reception times are recorded by the local clocks of the anchors and target $u$. We make a fundamental assumption that, due to the short propagation time duration\footnote{The range of 300 m corresponds to an approximate time delay of 1 $\mu$s, which is significantly shorter than the time slot.}, 
the displacement of target $u$ during the airtime is negligible, allowing the propagation time to be considered constant.
$\tau_{ui} (T_{ui}^{(m)})$ and $\tau_{uj} (T_{uj}^{(n)})$ represent the propagation times from anchor $i$ and anchor $j$ to target $u$ at $T_{ui}^{(m)}$ and $T_{uj}^{(n)}$, respectively. The clock offsets of target $u$ at different times are indicated in the figure, as well.

To obtain the TDOA at $T_{ui}^{(m)}$, it is necessary to determine the propagation times from anchor $i$ and anchor $j$ to target $u$ at the same instant (and same position), e.g., $\tau_{ui}(T_{ui}^{(m)})$ and $\tau_{uj}(T_{ui}^{(m)})$. However, since target $u$ is asynchronous and mobile, both the clock offset of target $u$ and the propagation time $\tau_{uj}$ vary over time. Therefore, first, we need to compute $\tau_{uj}(T_{ui}^{(m)})$ based on $\tau_{uj}(T_{uj}^{(m)})$, the clock offset and the target displacement. Then, $\tau_{u;ij}(T_{ui}^{(m)})$ can be calculated as follows: 
\begin{equation}
\begin{split}
    \tau_{u;ij}(T_{ui}^{(m)}) =& \tau_{ui}(T_{ui}^{(m)}) - \tau_{uj}(T_{ui}^{(m)})  \\
    =& \tau_{ui}(T_{ui}^{(m)}) - \left( \tau_{uj}(T_{uj}^{(n)}) - \Delta \tau_{uj} \right) \\
    =& \left( T_{ui}^{(m)}- T_{i}^{(m)}-\phi_u(T_{ui}^{(m)}) \right) - \\ &\left(T_{uj}^{(n)}- T_{j}^{(n)}-\phi_u(T_{uj}^{(n)}) \right) + \Delta \tau_{uj} \\
    =&  ( T_{ui}^{(m)}- T_{i}^{(m)} ) - (T_{uj}^{(n)}- T_{j}^{(n)}) + \\
    &\left( \phi_u(T_{uj}^{(n)}) - \phi_u(T_{ui}^{(m)}) \right)  + \Delta \tau_{uj}\\
    =& ( T_{ui}^{(m)}- T_{i}^{(m)} ) - (T_{uj}^{(n)}- T_{j}^{(n)}) + \Delta \phi_u + \Delta \tau_{uj}
\end{split}
\label{eq3}
\end{equation}
where $\Delta \phi_u$ and $\Delta \tau_{uj}$ represent the variations in the clock offset $\phi_u$ and the propagation time $\tau_{uj}$, which must be compensated.

There are simplified special cases for \eqref{eq3}. i) Target $u$ is stationary and has a frequency synchronized clock compared to the system clock ($\omega_u=1$), where the propagation time $\tau_{uj}$ and the clock offset $\phi_u$ remains constant, yielding $\Delta \tau_{uj}=\Delta \phi_u=0$. ii) Target $u$ is stationary but has an asynchronous clock, where $\Delta \tau_{uj} = 0$ still holds, while $\Delta \phi_{u} \neq 0$. In these two cases, TDOA estimation can be achieved by modeling the asynchronous clock in \eqref{eq2} as
\begin{equation}
    \Delta \phi_u = (1-\alpha_u)(T_{uj}^{(n)} - T_{ui}^{(m)})
    \label{eq4}
\end{equation}
where the drift calibration parameter $\alpha_u$ is associated with the clock drift, which can be estimated using methods outlined in prior studies \cite{dotlic2018ranging,puatru2023flextdoa,zhang2021signal}.

However, we aim to address a more general case involving a mobile target with an asynchronous clock, where both $\Delta \phi_u \neq 0$ and $\Delta \tau_{uj} \neq 0$. While the challenge of asynchronous clocks has been discussed previously, target mobility introduces additional complexity. Specifically, displacement during the reception interval causes measurements taken at different times to correspond to different locations. Since accurate TDOA estimation requires measurements tied to the same position, this spatial mismatch complicates the process, making the estimation of $\tau_{u;ij}$ a difficult problem. In other words, the significant challenge in the general case is the combined effect of the asynchronous clock and target mobility. While the former (asynchronous clock) has been extensively studied, the latter (target mobility) and the combined of the two have received little attention.

Similarly, if TDOA at $T_{uj}^{(n)}$ is also required, compensation for $\Delta \tau_{ui}$ becomes necessary. Therefore, instead of separately modeling $\Delta \tau_{ui}$ and $\Delta \tau_{uj}$, we propose to directly model the time-varying TDOA $\tau_{u;ij}$. We believe it is more effective and will introduce more details in Section \ref{PTDOA}.

\section{Parameterized TDOA} \label{PTDOA}

To localize the mobile target, concurrent TDOA measurements are typically required. However, this is impractical in a TDBPS system, where signals from different anchors are sequentially received by the target at various positions. In this section, we propose a method for TDOA estimation using sequential measurements, which allows the derivation of concurrent TDOA estimates by assigning the same time for the TDOA estimation of multiple anchor pairs.

Our new method adopts a parameterized approach, referred to as parameterized TDOA, or P-TDOA. To address the two primary challenges in TDOA estimation within a TDBPS, as outlined in Section \ref{challenge}, we first eliminate the asynchronous clock parameters, resulting in equations that solely involve the unknown propagation times. Next, to account for the mobility of the target, we introduce a polynomial function to model the time-varying propagation times, rather than directly modeling the target's motion. The original equations, which contain the unknown propagation times, are then transformed into equations involving the unknown parameters of the propagation time models. Under certain conditions, these equations can be further simplified to involve only the differences between the parameters of the propagation time models, which we refer to as the parameters of the polynomial TDOA models. Subsequently, based on the measurement protocol of our problem settings, we can collect data across multiple frames to form a system of linear equations in terms of the TDOA model parameters, which are then estimated using a WLS approach. Finally, to ensure the validity and well-posedness of the aforementioned WLS problem, we propose a strategy for constructing the equations. These steps are elaborated in the following subsections. Without loss of generality, we discuss a pair of anchors $(i, j)$ for illustration, where $i < j$.

\subsection{Clock Parameter Elimination} \label{clockEliminate}

The equation for the propagation time of the $m$-th signal transmitted by anchor $i$ in the system time coordinate is written as
\begin{equation}
T_i^{(m)} +\tau_{ui}(T_{ui}^{(m)})= \mathcal{H}(T_{ui}^{(m)}) =  \alpha_{u}T_{ui}^{(m)}+\beta_{u}
\label{eq5}
\end{equation}
For simplicity, we abbreviate the propagation time $\tau_{ui}(T_{ui}^{(m)})$ as $\tau_{ui}^{(m)}$ in subsequent illustrations, with similar abbreviations applied to other terms. The propagation time is coupled with the unknown clock calibration parameters of target $u$. To eliminate the offset calibration parameter $\beta_{u}$, we can write another equation for the propagation time of the $n$-th signal transmitted by anchor $j$ as
\begin{equation}
T_j^{(n)}+\tau_{uj}^{(n)}=\alpha_{u}T_{uj}^{(n)}+\beta_{u}
\label{eq6}
\end{equation}
Subtracting \eqref{eq5} from \eqref{eq6} results in
\begin{equation}
    (T_i^{(m)} - T_j^{(n)})+(\tau_{ui}^{(m)}-\tau_{uj}^{(n)})
=\alpha_{u}(T_{ui}^{(m)}-T_{uj}^{(n)})
\label{eq7}
\end{equation}
To eliminate the drift calibration parameter $\alpha_{u}$, we write another equation for the $p$-th signal from anchor $i$ and the $q$-th signal from anchor $j$, following the pattern of \eqref{eq7}, as
\begin{equation}
    (T_i^{(p)} - T_j^{(q)})+(\tau_{ui}^{(p)}-\tau_{uj}^{(q)})
=\alpha_{u}(T_{ui}^{(p)}-T_{uj}^{(q)})
\label{eq8}
\end{equation}
Dividing \eqref{eq7} by \eqref{eq8} yields a simplified relationship without $\alpha_{u}$ as given by
\begin{equation}
\frac{ (T_i^{(m)}  - T_j^{(n)}) +(\tau_{ui}^{(m)}-\tau_{uj}^{(n)}) }
{ (T_i^{(p)} - T_j^{(q)}) +(\tau_{ui}^{(p)}-\tau_{uj}^{(q)}) } 
=\frac{T_{ui}^{(m)}-T_{uj}^{(n)}}{T_{ui}^{(p)}-T_{uj}^{(q)}}
\label{eq9}
\end{equation}
We rearrange \eqref{eq9} to yield \eqref{eq10}, which is a linear function of the unknown propagation times, without involving any unknown clock parameters. The underbraced term $b_{u;ij}^{(m,n,p,q)}$ is a constant, where the superscript $(m,n,p,q)$ indicates its association with the $(m,n,p,q)$-th messages and this definition similarly applies to the other terms.
\fontsize{9}{10}
{
\begin{equation}
\begin{aligned}
    (T_{ui}^{(p)}-T_{uj}^{(q)})(\tau_{ui}^{(m)}-\tau_{uj}^{(n)})-(T_{ui}^{(m)}-T_{uj}^{(n)})(\tau_{ui}^{(p)}-\tau_{uj}^{(q)}) -  \\
\underbrace{\left[(T_i^{(p)} - T_j^{(q)})(T_{ui}^{(m)}-T_{uj}^{(n)}) - (T_i^{(m)} - T_j^{(n)})(T_{ui}^{(p)}-T_{uj}^{(q)})\right]}_{b_{u;ij}^{(m,n,p,q)}}  = 0
\end{aligned}
\label{eq10}
\end{equation}
}
\subsection{Polynomial TDOA Model} \label{TDOAmodel}
When target $u$ is moving, the distances or propagation times between it and different anchors become time-varying. To model the time-varying propagation times $\tau_{ui}$ and $\tau_{uj}$ between target $u$ and anchor $i$ and anchor $j$, we introduce a polynomial function with $L$ parameters, corresponding to an $(L-1)$-th order polynomial, expressed in terms of target $u$'s local time $T_u$\cite{rajan2015joint}.
\begin{equation}
\begin{aligned}
     \tau_{ui} (T_u) &=  \gamma_{0,ui}  + \ldots + \gamma_{L-1,ui} T_u^{L-1} + O(T_u^L)  \\
    &= \boldsymbol{\nu}_{T_u}^T \boldsymbol{\gamma}_{ui} + O(T_u^L) \\
    \tau_{uj} (T_u) &=  \gamma_{0,uj}  + \ldots + \gamma_{L-1,uj} T_u^{L-1} + O(T_u^L)  \\
    &= \boldsymbol{\nu}_{T_u}^T \boldsymbol{\gamma}_{uj} + O(T_u^L)
\end{aligned}
\label{eq11}
\end{equation}
 where
 \begin{gather}
    \boldsymbol{\gamma}_{ui}= [\gamma_{0,ui}, \ldots, \gamma_{L-1,ui} ]^T \in \mathbb{R}^{L\times1} \notag \\
    \boldsymbol{\gamma}_{uj}= [\gamma_{0,uj}, \ldots, \gamma_{L-1,uj} ]^T \in \mathbb{R}^{L\times1} \notag \\
    \boldsymbol{\nu}_{T_u}=[T_u^0,\ldots,T_u^{L-1}]^T \in \mathbb{R}^{L\times1} \notag
\end{gather}
$\boldsymbol{\gamma}_{ui}$ and $\boldsymbol{\gamma}_{uj}$ denote the model parameters of the propagation times from anchor $i$ and anchor $j$ to target $u$, respectively, which have incorporated the clock discrepancy of target $u$. The vector $\boldsymbol{\nu}_{T_u}$ consists of powers of target $u$'s local time $T_u$, ranging from order zero to order $L-1$. The term $O({T_u}^L)$ represents the higher-order error. In most scenarios, 
$L=2$, corresponding to a linear model that assumes the propagation time varies with constant speed, or $L=3$, corresponding to a second-order model that assumes the propagation time changes with constant acceleration, is sufficient to capture the variations of the time-varying propagation time over a short period, as shown in previous studies \cite{rajan2015joint}. Consequently, we neglect the higher-order errors in subsequent illustrations and restrict our discussions to cases where $L\leq3$.

By substituting the polynomial propagation time model in \eqref{eq11} into \eqref{eq10}, we transform the linear equation involving the unknown propagation times into one in terms of the unknown model parameters of the propagation times as given by
\begin{gather}
     \sum_{l=0}^{L-1}  \left( a_{l,ui}^{(m,n,p,q)} \gamma_{l,ui}
      -   a_{l,uj}^{(m,n,p,q)}\gamma_{l,uj} \right)
       - b_{u;ij}^{(m,n,p,q)} =0   \label{eq12} \\
       a_{l,ui}^{(m,n,p,q)} =  {T_{ui}^{(m)}}^{l}(T_{ui}^{(p)}-T_{uj}^{(q)})-{T_{ui}^{(p)}}^{l}(T_{ui}^{(m)}-T_{uj}^{(n)})  \notag \\
       a_{l,uj}^{(m,n,p,q)} =  {T_{uj}^{(n)}}^{l}(T_{ui}^{(p)}-T_{uj}^{(q)})-{T_{uj}^{(q)}}^{l}(T_{ui}^{(m)}-T_{uj}^{(n)})  \notag
\end{gather}
where $a_{l,ui}^{(m,n,p,q)}$ and $a_{l,uj}^{(m,n,p,q)}$ represent the coefficients of the $l$-th parameter $(l=0,1\ldots,L-1)$ in $\boldsymbol{\gamma}_{ui}$ and $\boldsymbol{\gamma}_{uj}$, respectively.

Different from reference \cite{rajan2015joint} where two-way communication is used to estimate the propagation time model parameters, in a TDBPS, we can only rely on one-way communication, based on which those parameters are intuitively unsolvable. However, through the following analysis, we demonstrate a novel method to solve the \textit{differences} of the propagation time model parameters instead of the original parameters themselves. Upon closer examination of the terms containing $\gamma_{l,ui}$ and $\gamma_{l,uj}$ in \eqref{eq12}, we rewrite them as follows:
\begin{equation}
\begin{split}
     &{a_{l,ui}^{(m,n,p,q)}\gamma_{l,ui}-a_{l,uj}^{(m,n,p,q)}\gamma_{l,uj}} \\
     & = a_{l,ui}^{(m,n,p,q)}(\gamma_{l,ui}-\gamma_{l,uj})
     + (a_{l,ui}^{(m,n,p,q)}-a_{l,uj}^{(m,n,p,q)})\gamma_{l,uj} \\
    & = a_{l,ui}^{(m,n,p,q)} \gamma_{l,u;ij} + \Delta a_{l,u;ij}^{(m,n,p,q)} \gamma_{l,uj} 
\end{split}
\label{eq13}
\end{equation}
where $\gamma_{l,u;ij} = \gamma_{l,ui}-\gamma_{l,uj}$ represents the difference of the parameters of the propagation time models, and
\begin{equation}
\begin{split}
\Delta a_{l,u;ij}^{(m,n,p,q)} = & a_{l,ui}^{(m,n,p,q)} - a_{l,uj}^{(m,n,p,q)}\\
= &({T_{ui}^{(m)}}^{l} - {T_{uj}^{(n)}}^{l})(T_{ui}^{(p)}-T_{uj}^{(q)}) \\
&-({T_{ui}^{(p)}}^{l} - {T_{uj}^{(q)}}^{l})(T_{ui}^{(m)}-T_{uj}^{(n)})
\label{eq14}
\end{split}
\end{equation}
represents the difference between the coefficients of the propagation time model parameters.

We consider the case
\begin{equation}
\Delta a_{l,u;ij}^{(m,n,p,q)} =  0
\label{eq15}
\end{equation}
and will develop a equation construction strategy to let it hold in Section \ref{STDS}. Then, $\gamma_{l,uj}$ is eliminated in \eqref{eq13}, thereby reducing \eqref{eq12} to an equation that only involves the differences of the parameters of the propagation time models. Subsequently, based on \eqref{eq12}, \eqref{eq13} and \eqref{eq15}, we derive \eqref{eq16}, where the coefficient $a_{l,ui}^{(m,n,p,q)}$ of $\gamma_{l,u;ij}$ in \eqref{eq13} is renamed as $a_{l,u;ij}^{(m,n,p,q)}$ to align with the parameter, as given by
\begin{gather}
 \sum_{l=0}^{L-1}  a_{l,u;ij}^{(m,n,p,q)} \gamma_{l,u;ij}  - b_{u;ij}^{(m,n,p,q)} =0 \label{eq16} \\
a_{l,u;ij}^{(m,n,p,q)} = {T_{ui}^{(m)}}^{l}(T_{ui}^{(p)}-T_{uj}^{(q)})-{T_{ui}^{(p)}}^{l}(T_{ui}^{(m)}-T_{uj}^{(n)})  \notag
\end{gather}
We denote
\begin{gather}
\boldsymbol{\gamma}_{u;ij} = [\gamma_{0,u;ij}, \ldots, \gamma_{L-1,u;ij} ]^T \in \mathbb{R}^{L\times1} \notag 
\end{gather}
as the parameters of a polynomial TDOA model, which approximates the time-varying TDOA as a polynomial function of time. Under this model, the TDOA between target $u$ and anchor pair $(i,j)$ at target $u$'s local time $T_u$ is expressed as follows:
\begin{equation}
    \tau_{u;ij} (T_u)  = \boldsymbol{\nu}_{T_u}^T \boldsymbol{\gamma}_{u;ij}
\label{eq17}
\end{equation}
In this way, once $\boldsymbol{ \gamma}_{u;ij}$ is estimated, the TDOA estimates can be derived using \eqref{eq17}.

\subsection{Mobile Weighted Least Squares} \label{MWLS}

We substitute the noise-free transmission and reception times in \eqref{eq16} with the noisy time measurements, and come to
\begin{equation}
  \hat{b}_{u;ij}^{(m,n,p,q)}={\boldsymbol{\hat{a}}_{u;ij}^{(m,n,p,q) 
 \ T}}\boldsymbol{\gamma}_{u;ij}+\eta_{u;ij}^{(m,n,p,q)}
  \label{eq18}
\end{equation}
where $\eta_{u;ij}^{(m,n,p,q)}$ represents the combined noise, including the noise in transmission and reception time measurements, the hat symbols above $\hat{b}_{u;ij}^{(m,n,p,q)}$ and ${\boldsymbol{\hat{a}}_{u;ij}^{(m,n,p,q)}}$ indicate that they are estimates contaminated by noise, the explicit expression of $\hat{b}_{u;ij}^{(m,n,p,q)}$, $\boldsymbol{\hat{a}}_{u;ij}^{(m,n,p,q)}$ and ${\eta}_{u;ij}^{(m,n,p,q)}$ are given in \eqref{eq19}, \eqref{eq20}, \eqref{eq21}, respectively. 
\fontsize{9}{10}
{
\begin{equation}
 \hat{b}_{u;ij}^{(m,n,p,q)} = (\hat{T}_i^{(p)} - \hat{T}_j^{(q)} )(\hat{T}_{ui}^{(m)}-\hat{T}_{uj}^{(n)})
- (\hat{T}_i^{(m)} - \hat{T}_j^{(n)})(\hat{T}_{ui}^{(p)}-\hat{T}_{uj}^{(q)})
\label{eq19}
\end{equation}
}
\begin{equation}
\begin{aligned}
    \boldsymbol{\hat{a}}_{u;ij}^{(m,n,p,q)} &= [ { \hat{a}_{0,u;ij}^{(m,n,p,q)}}, \ldots,{\hat{a}_{l,u;ij}^{(m,n,p,q)}}, \ldots, {\hat{a}_{L-1,u;ij}^{(m,n,p,q)}} ]^T  \\
\hat{a}_{l,u;ij}^{(m,n,p,q)} &=  {\hat{T}_{ui}^{(m) l}}(\hat{T}_{ui}^{(p)}-\hat{T}_{uj}^{(q)}) - {\hat{T}_{ui}^{(p) l}}(\hat{T}_{ui}^{(m)}-\hat{T}_{uj}^{(n)})  
\end{aligned}
 \label{eq20}
\end{equation}
\begin{equation}
    \begin{aligned}
    {\eta}_{u;ij}^{(m,n,p,q)} = \ &{\eta}_{u;ij(v)}^{(m,n,p,q)} + {\eta}_{u;ij(w)}^{(m,n,p,q)}    \\
     {\eta}_{u;ij(v)}^{(m,n,p,q)} = \ &(\hat{T}_{ui}^{(m)} - \hat{T}_{uj}^{(n)}) ( v_i^{(p)} - v_j^{(q)}) - \\
     &(\hat{T}_{ui}^{(p)} - \hat{T}_{uj}^{(q)}) ( v_i^{(m)} - v_j^{(n)})  \\
      {\eta}_{u;ij(w)}^{(m,n,p,q)} = \ &(\hat{T}_i^{(p)} - \hat{T}_j^{(q)})(w_{ui}^{(m)} - w_{uj}^{(n)}) - \\
    &(\hat{T}_i^{(m)} - \hat{T}_j^{(n)})(w_{ui}^{(p)} - w_{uj}^{(q)})
\end{aligned}
\label{eq21}
\end{equation}
The terms $\eta_{u;ij(v)}^{(m,n,p,q)}$ and $\eta_{u;ij(w)}^{(m,n,p,q)}$ represent the noise contributions from transmission time measurements and reception time measurements, respectively.

In the problem settings introduced in Section \ref{problemFormulation}, each anchor broadcasts once per frame, and we define $N_f$ frames as a solution period. To construct an equation similar to \eqref{eq18}, four distinct received signals are required: two signals each transmitted from anchor $i$ and anchor $j$, respectively, meaning that data from at least two frames are needed. Consequently, if we consider all the recorded times within a solution period, at most $N_f-1$ independent equations can be constructed. Suppose that we obtain $M$ equations in terms of $\boldsymbol{\gamma}_{u;ij}\in \mathbb{R}^{L\times1}$. To ensure a unique solution, it is required that $M\geq L$.\footnote{As previously mentioned, the number of available independent equations $M\leq N_f-1$, which implies that $N_f \geq L+1$. Thus, the minimum requirement for a solution period is $L+1$ frames.} By combining the $M$ equations, we formulate a system of linear equations
\begin{equation}
\boldsymbol{\hat{b}}_{u;ij}=\boldsymbol{\hat{A}}_{u;ij}\boldsymbol{\gamma}_{u;ij}+\boldsymbol{\eta}_{u;ij}
\label{eq22}
\end{equation}

where
\begin{equation}
  \boldsymbol{\hat{b}}_{u;ij}=[\hat{b}_{u;ij}^{(1)},\hat{b}_{u;ij}^{(2)},\ldots,\hat{b}_{u;ij}^{(M)}]^T
  \label{eq23}
\end{equation}
\begin{equation}
  \boldsymbol{\hat{A}}_{u;ij}=[\boldsymbol{\hat{a}}_{u;ij}^{(1)}, \boldsymbol{\hat{a}}_{u;ij}^{(2)}, \ldots, \boldsymbol{\hat{a}}_{u;ij}^{(M)}]^T
  \label{eq24}
\end{equation}
\begin{equation}
  \boldsymbol{\eta}_{u;ij}=[\eta_{u;ij}^{(1)},\eta_{u;ij}^{(2)},\ldots,\eta_{u;ij}^{(M)}]^T
  \label{eq25}
\end{equation}
and the superscripts of $\hat{b}_{u;ij}$, $\boldsymbol{\hat{a}}_{u;ij}$ and $\eta_{u;ij}$ are the equation indices, each representing a unique frame combination of $(m,n,p,q)$ in \eqref{eq18}.

With the system of linear equations in \eqref{eq22}, we estimate $\boldsymbol{\gamma}_{u;ij}$ by WLS as
\begin{gather}
     \boldsymbol{\hat{\gamma}}_{u;ij}=(\boldsymbol{\hat{A}}_{u;ij}^T \boldsymbol{\Sigma}_{\boldsymbol{\eta}_{u;ij}}^{-1} \boldsymbol{\hat{A}}_{u;ij})^{-1} \boldsymbol{\hat{A}}_{u;ij}^T \boldsymbol{\Sigma}_{\boldsymbol{\eta}_{u;ij}}^{-1} \boldsymbol{\hat{b}}_{u;ij} 
     \label{eq26}
     \\
     \boldsymbol{\Sigma}_{ \boldsymbol{\gamma}_{u;ij} } = (\boldsymbol{\hat{A}}_{u;ij}^T \boldsymbol{\Sigma}_{\boldsymbol{\eta}_{u;ij}}^{-1} \boldsymbol{\hat{A}}_{u;ij})^{-1} 
    \label{eq27}
\end{gather}
where $\boldsymbol{\Sigma}_{\boldsymbol{\boldsymbol{\eta}_{u;ij}}}$ and $\boldsymbol{\Sigma}_{ \boldsymbol{\gamma}_{u;ij} }$ represent the covariance matrices of the noise vector $\boldsymbol{\eta}_{u;ij}$ and the estimated TDOA model parameter vector $ \boldsymbol{\hat{\gamma}}_{u;ij}$, respectively. The matrix $\boldsymbol{\Sigma}_{\boldsymbol{\boldsymbol{\eta}_{u;ij}}}$ is determined by the measurement protocol and the specific equation construction strategy employed. A detailed example for our problem settings will be provided in Section \ref{theoreticalMSE}.

After obtaining $\boldsymbol{\hat{\gamma}}_{u;ij}$, the TDOA estimate $\tau_{u;ij}$ at $T_u$ can be derived by \eqref{eq17} as follows:
\begin{equation}
    \hat{\tau}_{u;ij}(T_u)= \boldsymbol{\nu}_{T_u}^T \boldsymbol{\hat{\gamma}}_{u;ij}
    \label{eq28}
\end{equation}
and the variance of the TDOA estimate $\hat{\tau}_{u;ij}$ is given by
\begin{equation}
    \sigma_{\tau_{u;ij}}^2 =  \boldsymbol{\nu}_{T_u}^T \boldsymbol{\Sigma}_{ \boldsymbol{\gamma}_{u;ij} } \boldsymbol{\nu}_{T_u}
    \label{eq29}
\end{equation}

At this stage, by aligning the estimated TDOA models of multiple anchor pairs to the same instant, concurrent TDOA estimates can be derived, which can then be directly used for localization using classical TDOA-based methods. Since this algorithm is designed for mobile targets, we refer to it as mobile weighted least squares, namely MWLS.

\subsection{Successive Time Difference Strategy} \label{STDS}

In this subsection, we examine the conditions under which the assumption $\Delta a_{l,u;ij}^{(m,n,p,q)}=0$ holds. We organize the discussion based on different values of $l$.

\subsubsection{$l=0,1$} \quad \\
\indent It is evident from \eqref{eq14} that $\Delta a_{0,u;ij}^{(m,n,p,q)}=0$ and $\Delta a_{1,u;ij}^{(m,n,p,q)}=0$ hold for any assignment of $(m,n,p,q)$:
\begin{equation}
\begin{aligned}
    \Delta a_{0,u;ij}^{(m,n,p,q)} = & (1-1)(T_{ui}^{(p)}-T_{uj}^{(q)})-(1-1)(T_{ui}^{(m)}-T_{uj}^{(n)}) = 0 \\
    \Delta a_{1,u;ij}^{(m,n,p,q)} = & (T_{ui}^{(m)}-T_{uj}^{(n)})(T_{ui}^{(p)}-T_{uj}^{(q)})- \\
    &(T_{ui}^{(p)}-T_{uj}^{(q)})(T_{ui}^{(m)}-T_{uj}^{(n)}) = 0
\label{eq30}
\end{aligned}
\end{equation}

\subsubsection{$l=2$} \quad \\
\indent $\Delta a_{2,ij}^{(m,n,p,q)}$ is given by
\begin{equation}
\begin{split}
     \Delta a_{2,u;ij}^{(m,n,p,q)}= &({T_{ui}^{(m)}} - {T_{uj}^{(n)}})(T_{ui}^{(p)}-T_{uj}^{(q)}) \cdot \\
     &({T_{ui}^{(m)}} + {T_{uj}^{(n)}} - T_{ui}^{(p)}- T_{uj}^{(q)})
    \label{eq31}
\end{split}
\end{equation}
Under the measurement protocol in our problem settings, the following condition holds:
\begin{equation}
\begin{split}
    &{T_{ui}^{(m)}} + {T_{uj}^{(n)}} - T_{ui}^{(p)}- T_{uj}^{(q)} \\
    = &(m-p)T_f + (n-q)T_f \\
    = &(m+n-p-q)T_f
\end{split}
\label{eq32}
\end{equation}
where $T_f$ is the duration of one frame. It implies that if 
\begin{equation}
    m+n-p-q=0
    \label{eq33}
\end{equation}
holds for certain assignments of $(m,n,p,q)$, then $\Delta a_{2,u;ij}^{(m,n,p,q)}=0$ is satisfied.

\begin{figure}
    \centering    \includegraphics[width=\linewidth]{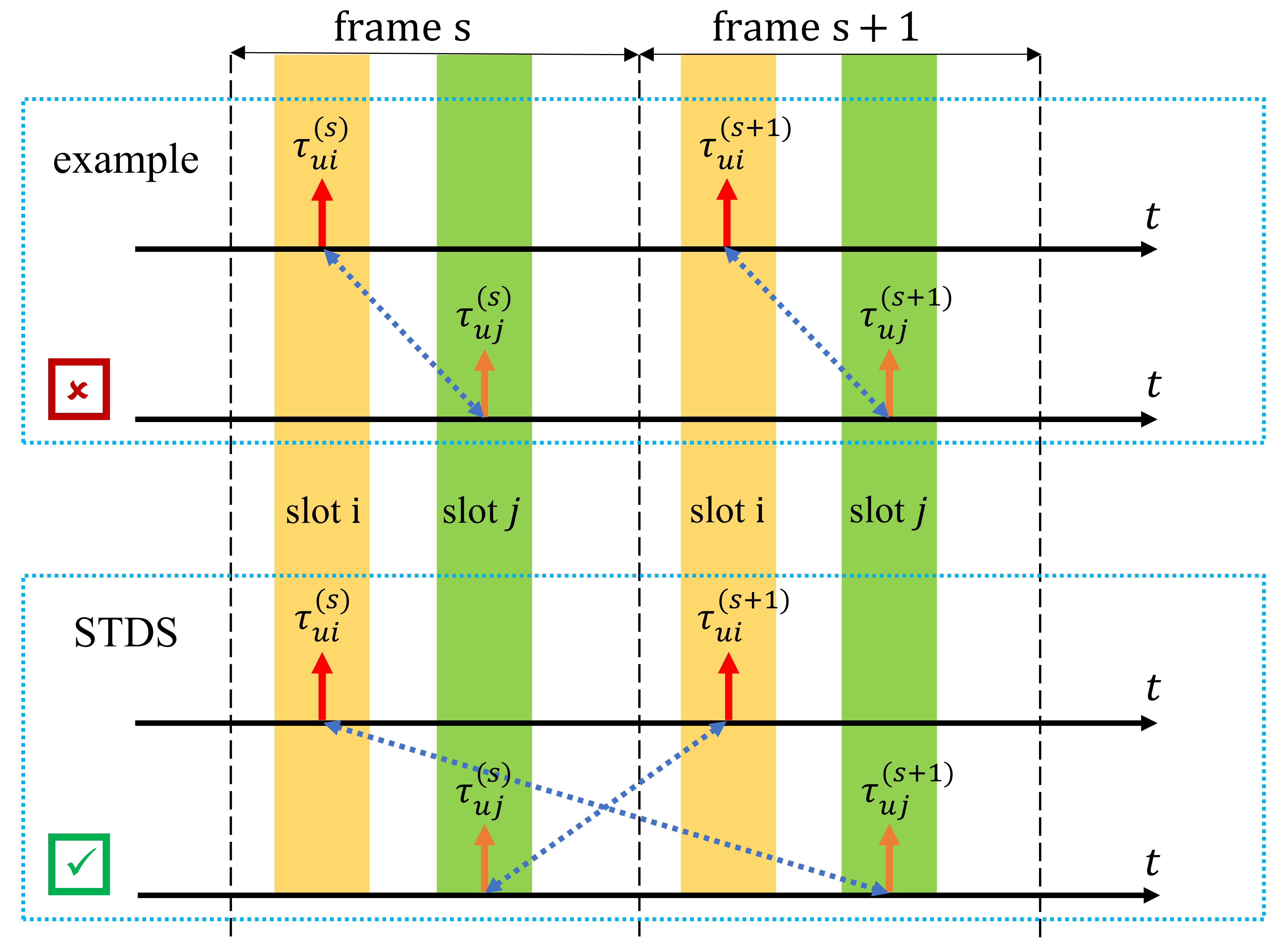}
    \caption{
    Equation construction strategy by dedicated frame selection. The red arrows represent the propagation times from anchor $i$ to target $u$, while the orange arrows represent the propagation times from anchor $j$ to target $u$. The blue dashed two-headed arrows represent the combination of these values into a single equation through their differences, as shown in \eqref{eq7} and \eqref{eq8}. The upper part demonstrates an example strategy, which uses propagation time differences within the same frame and has a rank-deficiency issue. The lower part shows our proposed STDS, which utilizes propagation time differences from successive frames, and avoids the rank-deficiency issue.
    }
    \label{fig4}
\end{figure}

Furthermore, it is important to note that not all assignments that satisfy \eqref{eq32} are acceptable. We must ensure that the resulting least squares problem is well-posed. For instance, as illustrated in the upper part of Fig. \ref{fig4}, if $m=n=s,p=q=s+1$ ($s=1,\ldots,N_f-1$) is assigned to construct equations in \eqref{eq22}, the coefficients $a_{l,u;ij}^{(m,n,p,q)}$ for $l=0,1$ become
\begin{equation}
\begin{aligned}
     a_{0,u;ij}^{(s,s,s+1,s+1)}&=(T_{ui}^{(s+1)}-T_{uj}^{(s+1)})- (T_{ui}^{(s)}-T_{uj}^{(s)}) \\
    &\approx 0 \\
    a_{1,u;ij}^{(s,s,s+1,s+1)} &=T_{ui}^{(s)}(T_{ui}^{(s+1)}-T_{uj}^{(s+1)})- T_{ui}^{(s+1)}(T_{ui}^{(s)}-T_{uj}^{(s)}) \\
    & \approx (j-i)T_sT_f
\end{aligned}
\label{eq34}
\end{equation}
where $T_s$ is the duration of one slot as shown by Fig. \ref{fig2}. In this case, $a_{0,u;ij}^{(s,s,s+1,s+1)}\approx 0$ and $a_{1,u;ij}^{(s,s,s+1,s+1)}$ is approximately constant for all $s$. Consequently, 
two columns of the coefficient matrix $\boldsymbol{\hat{A}}_{u;ij}$ in \eqref{eq24} become nearly fully correlated, resulting in $\boldsymbol{\hat{A}}_{u;ij}$ being rank-deficient, which renders the problem ill-posed.

In contrast, we propose an equation construction strategy where
\begin{equation}
m=q=s,\; p=n=s+1,\; s=1,\cdots,N_f-1
\label{eq35}
\end{equation}
is assigned to construct equations in \eqref{eq22}, as illustrated in the subpart of Fig. \ref{fig4}. The coefficients $a_{l,u;ij}^{(m,n,p,q)}$ for $l=0,1,2$ become
\begin{equation}
\begin{aligned}
     a_{0,u;ij}^{(s,s+1,s+1,s)} &=(T_{ui}^{(s+1)}-T_{uj}^{(s)})- (T_{ui}^{(s)}-T_{uj}^{(s+1)}) \\
     &\approx 2T_f \\
    a_{1,u;ij}^{(s,s+1,s+1,s)} &=T_{ui}^{(s)}(T_{ui}^{(s+1)}-T_{uj}^{(s)})- T_{ui}^{(s+1)}(T_{ui}^{(s)}-T_{uj}^{(s+1)}) \\
    & \approx T_f(T_{ui}^{(s+1)}+T_{uj}^{(s)}) \\
    a_{2,u;ij}^{(s,s+1,s+1,s)} &={T_{ui}^{(s)}}^2(T_{ui}^{(s+1)}-T_{uj}^{(s)})- {T_{ui}^{(s+1)}}^2 (T_{ui}^{(s)}-T_{uj}^{(s+1)}) \\
    & \approx T_f(T_{ui}^{(s+1)}T_{uj}^{(s+1)}+T_{ui}^{(s)}T_{uj}^{(s)}) 
\end{aligned}
\label{eq36}
\end{equation}
According to \eqref{eq36}, it turns out that by applying our strategy, the columns of $\boldsymbol{\hat{A}}_{u;ij}$ in \eqref{eq24} become linearly uncorrelated, thus avoiding the issue encountered in the previous example and ensuring that the coefficient matrix $\boldsymbol{\hat{A}}_{u;ij}$ has full column rank, which ultimately leads to a well-posed problem. Since the equation construction strategy leverages propagation time differences from successive frames, as shown in \eqref{eq7} and \eqref{eq8}, we refer to it as the successive time difference strategy, namely STDS.

\section{Performance Analysis}
In this section, we analyze performance of our proposed method. Since higher-order errors have been neglected, the noise vector $\boldsymbol{\eta}_{u;ij}$ in \eqref{eq25} is considered zero-mean, which implies that the derived model parameter estimates and TDOA estimates are unbiased. Consequently, CRLB can be used to evaluate the performance of our proposed method.
However, sequential measurements are unsuitable for analyzing the CRLB for TDOA estimation, as they are acquired at different times and positions. Therefore, we analyze the case of concurrent measurements, where the target receives signals from multiple anchors simultaneously. It is important to note that actual TDOA measurements can only be obtained in frequency-division (FD) or code-division (CD) systems, such as GNSS, and are not feasible in practical TD systems. The CRLB derived here represents a theoretical limit that is hard to achieve with sequential measurements, serving as a reference lower bound for performance analysis.

\subsection{Cram\'er-Rao Lower Bound}\label{CRLB}

To derive the CRLB for TDOA estimation in concurrent measurement scenarios, we assume that the noise in transmission time and reception time measurements, introduced in \eqref{eq1}, is independent and identically distributed Gaussian noise with different variances, respectively, i.e., $v_{i}^{(m)} \sim \mathcal{N}(0,\sigma_t^2)$ and $w_{ui}^{(m)} \sim \mathcal{N}(0,\sigma_r^2)$ for all $i\in \mathcal{N}_a$ and $m=1,\ldots,N_f$. We take target $u$ and anchor pair $(i,j)$ in frame $m$ for example, where anchor $i$ and anchor $j$ broadcasts signals at $T_i^{(m)}$ and $T_j^{(m)}$ and target $u$ receives these signals at $T_{ui}^{(m)}$ and $T_{uj}^{(m)}$ ($T_{ui}^{(m)} =  T_{uj}^{(m)}$), respectively. Since the reception times are identical, the clock offsets of target $u$ are the same, denoted as $\phi_{u}^{(m)}$. The equations in terms of the propagation times can be written as 
\begin{equation}
\begin{aligned}
    T_i^{(m)} +\tau_{ui}^{(m)}  = T_{ui}^{(m)} -\phi_u^{(m)} \\
    T_j^{(m)} +\tau_{uj}^{(m)}  = T_{uj}^{(m)} -\phi_u^{(m)}
\end{aligned}
\label{eq37}
\end{equation}
Thus, the TDOA can be calculated by
\begin{equation}
\begin{split}
    \tau_{u;ij}^{(m)} &= \tau_{ui}^{(m)} - \tau_{uj}^{(m)} \\
    & = (T_{ui}^{(m)} - T_i^{(m)}) - (T_{uj}^{(m)} - T_j^{(m)})
\end{split}
\label{eq38}
\end{equation}
We incorporate the noise in recorded transmission and reception times to derive the actual TDOA measurement as follows
\begin{equation}
\begin{split}
    \hat{\tau}_{u;ij}^{(m)} =\ & (\hat{T}_{ui}^{(m)} - \hat{T}_i^{(m)}) - (\hat{T}_{uj}^{(m)} - \hat{T}_j^{(m)})\ + \\
    &(v_i^{(m)} - v_j^{(m)}) - (w_{ui}^{(m)} - w_{uj}^{(m)}) \\
    =\ & {\tau}_{u;ij}^{(m)} + \Delta {\tau}_{u;ij}^{(m)}
\end{split}
\label{eq39}
\end{equation}
where $\Delta {\tau}_{u;ij}^{(m)} = (v_i^{(m)} - v_j^{(m)}) - (w_{ui}^{(m)} - w_{uj}^{(m)})$ represents the noise in the TDOA measurement $\hat{\tau}_{u;ij}^{(m)}$. We denote $\sigma_n^2=2(\sigma_r^2+\sigma_t^2)$ as the variance of an independent TDOA measurement, i.e., $\Delta {\tau}_{u;ij}^{(m)} \sim \mathcal{N}(0, \sigma_n^2)$.

Assume that target $u$ obtains $N_f$ TDOA measurements from anchor pair $(i,j)$ in the whole solution period, denoted as $\hat{\boldsymbol{\tau}}_{u;ij}=[ \hat{\tau}_{u;ij}^{(1)},\hat{\tau}_{u;ij}^{(2)}, \ldots,\hat{\tau}_{u;ij}^{(N_f)}]^T$, at $\boldsymbol{T}_{u}=[T_u^{(1)},T_u^{(2)},\ldots,T_u^{(N_f)}]^T$. 
The covariance matrix of $\hat{\boldsymbol{\tau}}_{u;ij}$ is $\boldsymbol{\Sigma}_{\boldsymbol{\tau}_{u;ij}} = \sigma_n^2 \boldsymbol{I}_{N_f}$, which also represents the CRLB of TDOA estimation, denoted as
\begin{equation}
    \text{CRLB1}(\boldsymbol{\tau}_{u;ij}) = \sigma_n^2 \boldsymbol{I}_{N_f}
    \label{eq40}
\end{equation}
However, since we applies a polynomial model to the TDOA estimates in our new method, as shown in \eqref{eq17}, which effectively adds constraints to the estimates, it is necessary to consider the impact on the CRLB. Following the polynomial model in \eqref{eq17}, these TDOA estimates can be written as
\begin{equation}
\begin{split}
    \hat{\boldsymbol{\tau}}_{u;ij} = &[ \boldsymbol{\nu}_{T_u^{(1)}}^T \boldsymbol{\hat{\gamma}}_{u;ij},     \boldsymbol{\nu}_{T_u^{(2)}}^T \boldsymbol{\hat{\gamma}}_{u;ij}, \ldots, \boldsymbol{\nu}_{T_u^{(N_f)}}^T \boldsymbol{\hat{\gamma}}_{u;ij}]^T \\
     = &[ \boldsymbol{\nu}_{T_u^{(1)}}, \boldsymbol{\nu}_{T_u^{(2)}}, \ldots, \boldsymbol{\nu}_{T_u^{(N_f)}}   ]^T \boldsymbol{\hat{\gamma}}_{u;ij} \\
     = &[\boldsymbol{T}_u^{\odot 0}, \boldsymbol{T}_u^{\odot 1}, \ldots, \boldsymbol{T}_u^{\odot L-1}] \boldsymbol{\hat{\gamma}}_{u;ij} \\
     = & \boldsymbol{V}_u \boldsymbol{\hat{\gamma}}_{u;ij}
\end{split}
\label{eq41}
\end{equation}
where $\boldsymbol{V}_u \in \mathbb{R}^{N_f \times L}$ is defined as
\begin{equation}
    \boldsymbol{V}_u = \mathcal{V} ( \boldsymbol{T}_u ) =    [\boldsymbol{T}_u^{\odot 0}, \boldsymbol{T}_u^{\odot 1}, \ldots, \boldsymbol{T}_u^{\odot L-1}]
\label{eq42}
\end{equation}
Here, $\mathcal{V}$ represents an operator that converts a vector into the transpose of a Vandermonde matrix. 
Although $\boldsymbol{V}_u$ may be non-invertible, however, the Vandermonde matrix $\boldsymbol{V}_u$ is guaranteed to have full column rank if $N_f \geq L$. In this case, the matrix $\boldsymbol{V}_u$ has a unique left inverse, denoted as $\boldsymbol{V}_u^{\dagger}$ \cite{zhang2017matrix}, as given by
\begin{equation}
    \boldsymbol{V}_u^{\dagger}=\left(\boldsymbol{V}_u^T \boldsymbol{V}_u \right)^{-1} \boldsymbol{V}_u^T
    \label{eq43}
\end{equation}
We use $\boldsymbol{V}_u^{\dagger}$ to rewrite the relationship in \eqref{eq41} by
\begin{equation}
     \boldsymbol{\hat{\gamma}}_{u;ij} = \boldsymbol{V}_u^{\dagger} \hat{\boldsymbol{\tau}}_{u;ij}
     \label{eq44}
\end{equation}
The CRLB of $\boldsymbol{\gamma}_{u;ij}$ is derived as follows\cite{kay1993fundamentals}
\begin{equation}
\begin{split}
    \text{CRLB}(\boldsymbol{\gamma}_{u;ij}) &=\frac{\partial \boldsymbol{\gamma}_{u;ij}}{\partial  \boldsymbol{\tau}_{u;ij}}   \text{CRLB1}(\boldsymbol{\tau}_{u;ij})
    \frac{\partial \boldsymbol{\gamma}_{u;ij}}{\partial  \boldsymbol{\tau}_{u;ij}}^T \\
    &=  \boldsymbol{V}_u^{\dagger} \left(\sigma_n^2\boldsymbol{I}_{N_f} \right) {\boldsymbol{V}_u^{\dagger}}^T 
    = \sigma_n^2 \left( \boldsymbol{V}_u^T \boldsymbol{V}_u \right)^{-1}
\end{split}
\label{eq45}
\end{equation}
The CRLB for the TDOA estimates $\boldsymbol{\tau}_{u;ij}$ under the polynomial model is given by\footnote{Here we use CRLB1 and CRLB2 to distinguish between the CRLB for TDOA estimation without and with the polynomial TDOA model, respectively. A similar distinction applies to the CRLB for localization in the following sections.}:
\begin{equation}
\begin{split}
    \text{CRLB2}(\boldsymbol{\tau}_{u;ij}) & = \boldsymbol{V}_u \text{CRLB}(\boldsymbol{\gamma}_{u;ij})\boldsymbol{V}_u^T \\
    &= \sigma_n^2  \boldsymbol{V}_u \left( \boldsymbol{V}_u^T \boldsymbol{V}_u \right)^{-1}  \boldsymbol{V}_u^T
\end{split}
\label{eq46}
\end{equation}

\noindent\textbf{Remark}: \textit{The polynomial model reduces the CRLB for TDOA estimation, i.e.,} 
\begin{equation*}
    \text{CRLB1}(\boldsymbol{\tau}_{u;ij}) \succeq  \text{CRLB2}(\boldsymbol{\tau}_{u;ij})
\end{equation*}
\begin{proof}
We denote $\boldsymbol{P}=\boldsymbol{V}_u \left(\boldsymbol{V}_u^T \boldsymbol{V}_u \right)^{-1}  \boldsymbol{V}_u^T$, which is an orthogonal projection matrix satisfying $\boldsymbol{P}^2 = \boldsymbol{P}$ and $\boldsymbol{P}^T = \boldsymbol{P}$. 
For $\forall$ $ \boldsymbol{x} \in \mathbb{R}^{N_f \times 1}$, it holds that
\begin{equation}
\begin{aligned}
\boldsymbol{x}^T\boldsymbol{P}\boldsymbol{x}
=\boldsymbol{x}^T\boldsymbol{P}^2\boldsymbol{x}=\boldsymbol{x}^T\boldsymbol{P}^T\boldsymbol{P}\boldsymbol{x} =
{\left\lVert \boldsymbol{P}\boldsymbol{x} \right\rVert}^2 \leq  {\left\lVert \boldsymbol{x} \right\rVert}^2
\end{aligned}
\label{eq47}
\end{equation}
Next, we define $\boldsymbol{A} = \boldsymbol{I}_{N_f} - \boldsymbol{P}$, which satisfies
\begin{equation}
    \boldsymbol{x}^T \boldsymbol{A} \boldsymbol{x} 
    = \boldsymbol{x}^T
    \boldsymbol{x} - \boldsymbol{x}^T\boldsymbol{P}\boldsymbol{x} 
    = {\left\lVert \boldsymbol{x} \right\rVert}^2 - 
    {\left\lVert \boldsymbol{P}\boldsymbol{x} \right\rVert}^2 
    \geq 0
\label{eq48}
\end{equation}
This shows that $\boldsymbol{A}$ is a positive semidefinite matrix, i.e., $\boldsymbol{A} \succeq \boldsymbol{0}_{N_f}$. Therefore, the CRLBs for localization satisfy
\begin{equation}
    \text{CRLB1}(\boldsymbol{\tau}_{u;ij})- \text{CRLB2}(\boldsymbol{\tau}_{u;ij}) = 
    \sigma_n^2 \boldsymbol{A} \succeq \boldsymbol{0}_{N_f}
\label{eq49}
\end{equation}
which indicates that
\begin{equation}
    \text{CRLB1}(\boldsymbol{\tau}_{u;ij}) \succeq \text{CRLB2}(\boldsymbol{\tau}_{u;ij})
\label{eq50}
\end{equation}
\end{proof}
\subsection{Theoretical MSE}\label{theoreticalMSE}
We have derived the expressions for the covariances of the estimated parameters, $\boldsymbol{\hat{\gamma}}_{u;ij}$ and the estimated TDOA $\hat{\tau}_{u;ij}$ in \eqref{eq27} and \eqref{eq29}, respectively. However, these expressions contain the undetermined terms, $\boldsymbol{\hat{A}}_{u;ij}$ and $\boldsymbol{\Sigma}_{\boldsymbol{\eta}_{u;ij}}$. To evaluate the performance of the proposed P-TDOA, we analyze the theoretical MSE of $\boldsymbol{\hat{\gamma}}_{u;ij}$ and $\hat{\tau}_{u;ij}$ in our problem settings with the measurement protocol and STDS applied. The detailed derivation of $\boldsymbol{\Sigma}_{\boldsymbol{\eta}_{u;ij}}$ is illustrated in Appendix \ref{appendix1}, while the entries of the coefficients matrix $\boldsymbol{\hat{A}}_{u;ij}$ in \eqref{eq24} are all zero except for the following items:
\begin{equation}
\begin{split}
     \left[  \boldsymbol{\hat{A}}_{u;ij}  \right]_{s,l} =& \left(\hat{T}_{ui}^{(s)}\right)^{l - 1} \left(\hat{T}_{ui}^{(s+1)} - \hat{T}_{uj}^{(s)}\right) - \\
     & \left(\hat{T}_{ui}^{(s+1)}\right)^{l - 1} \left(\hat{T}_{ui}^{(s)} - \hat{T}_{uj}^{(s+1)}\right)
\end{split}
\label{eq51}
\end{equation}
where $s=1,2,\ldots,N_f-1$ and $l=0,1,\ldots,L-1$. To simplify the analysis, we introduce some approximations to the aforementioned matrices. As previously mentioned, the propagation times are negligible compared to the duration of a time slot or frame. By neglecting both the propagation times and measurement noise, it can be approximated that
\begin{equation}
\begin{aligned}
    \hat{T}_{i}^{(m)} - \hat{T}_{j}^{(n)} \approx
    \hat{T}_{ui}^{(m)} - \hat{T}_{uj}^{(n)} \approx (m-n)T_f +(i-j)T_s 
\end{aligned}
\label{eq52}
\end{equation}
Then $\boldsymbol{\Sigma}_{\boldsymbol{\eta}_{u;ij}}$ and $\boldsymbol{\hat{A}}_{u;ij}$ are rewritten as 
\begin{equation}
    \boldsymbol{\Sigma}_{\boldsymbol{\eta}_{u;ij}} = \sigma_n^2
    \left[ 
    \begin{matrix}
         (r_1^2 + r_2^2) &r_1r_2  &\cdots &0  \\
         r_1r_2 &(r_1^2 + r_2^2)  &\cdots &\vdots \\
         \vdots &\vdots &\ddots &r_1r_2  \\
         0  &\cdots &r_1r_2 &(r_1^2 + r_2^2)  \\
    \end{matrix}
    \right]    
 \label{eq53}
\end{equation}

\begin{gather}
    \boldsymbol{\hat{A}}_{u;ij} = \boldsymbol{B} \mathcal{V}(\boldsymbol{T}_{ui}) = \boldsymbol{B} \boldsymbol{V}_{ui} \notag \\
    \boldsymbol{B} =
    \left[ 
    \begin{matrix}
         r_1 &r_2 &0 &\cdots &0 &0  \\
         0 &r_1 &r_2 &\cdots &0 &0  \\
         \vdots &\vdots &\vdots &\ddots &\vdots &\vdots  \\
         0 &0 &0 &\cdots &r_1 &r_2  \\
    \end{matrix}
    \right]
\label{eq54}
\end{gather}
where $r_1$ and $r_2$ satisfy
\begin{equation}
    \begin{aligned}
        &r_1=T_f + (i-j) T_s \\
        &r_2=T_f - (i-j) T_s
    \end{aligned}
 \label{eq55}
\end{equation}
The operator $\mathcal{V}$ is defined behind \eqref{eq42}, and $\boldsymbol{T}_{ui} = [T_{ui}^{(1)}, \ldots, T_{ui}^{(N_f)} ]^T$ is the vector of target $u$'s reception times for signals from anchor $i$. 
By substituting \eqref{eq54} into \eqref{eq27} we obtain
\begin{equation}
\begin{split}
    \boldsymbol{\Sigma}_{\boldsymbol{\gamma}_{u;ij}} =& \left( 
    \boldsymbol{V}_{ui}^T \boldsymbol{B}^T \boldsymbol{\Sigma}_{\boldsymbol{\eta}_{u;ij}}^{-1} \boldsymbol{B} \boldsymbol{V}_{ui}
    \right)^{-1} \\
    =& \ \sigma_n^2 \left( 
    \boldsymbol{V}_{ui}^T \boldsymbol{D} \boldsymbol{V}_{ui}
    \right)^{-1}
\end{split}
\label{eq56}
\end{equation}
where $\boldsymbol{D}= \boldsymbol{B}^T \boldsymbol{\Sigma}_{\boldsymbol{\eta}_{u;ij}}^{-1} \boldsymbol{B} \cdot \sigma_n^2$. 

Since the inverse of the symmetric triple diagonal matrix in \eqref{eq53} has no straightforward analytical solution, simplifying $\boldsymbol{D}$ becomes extremely challenging. Therefore, we proceed by considering a condition that $T_f \gg \left\lvert i-j \right\rvert T_s$, which is explained later in this section. Then $\boldsymbol{D}$ can be approximated as
{\fontsize{9.5}{10}\selectfont 
\begin{equation}
\begin{split}
    \boldsymbol{D} & \approx
    \left[ 
    \begin{matrix}
         \frac{N_f-1}{N_f} &\frac{1}{N_f} &\cdots  &(-1)^{N_f} \frac{1}{N_f} \\
         \frac{1}{N_f} &\frac{N_f-1}{N_f} &\cdots &(-1)^{N_f+1} \frac{1}{N_f}\\
         \vdots &\vdots &\ddots & \vdots  \\
         (-1)^{N_f} \frac{1}{N_f} &(-1)^{N_f+1} \frac{1}{N_f} &\cdots &\frac{N_f-1}{N_f}
    \end{matrix}
    \right]  \\
    &= \boldsymbol{I}_{N_f} + \frac{1}{N_f}
    \left[ 
    \begin{matrix}
         -1 &1 &\cdots  &(-1)^{N_f} \\
         1 &-1 &\cdots &(-1)^{N_f+1}\\
         \vdots &\vdots &\ddots & \vdots  \\
         (-1)^{N_f} &(-1)^{N_f+1} &\cdots &(-1)^{2N_f-1}
    \end{matrix}
    \right] \\
    & = \boldsymbol{I}_{N_f} + \Delta \boldsymbol{D}
\end{split}
\label{eq57}
\end{equation}
} 
Substituting \eqref{eq57} into \eqref{eq56}, it yields that
\begin{equation}
\begin{split}
    \boldsymbol{\Sigma}_{\boldsymbol{\gamma}_{u;ij}} 
    &= \ \sigma_n^2 \left( \boldsymbol{V}_{ui}^T \boldsymbol{V}_{ui}+
    \boldsymbol{V}_{ui}^T  \Delta \boldsymbol{D} \boldsymbol{V}_{ui}
    \right)^{-1} \\
    &= \ \sigma_n^2 \left( \boldsymbol{F}_1 + \boldsymbol{F}_2 \right)^{-1}
\end{split}
\label{eq58}
\end{equation}
where $\boldsymbol{F}_1 = \boldsymbol{V}_{ui}^T \boldsymbol{V}_{ui}$, $\boldsymbol{F}_2 = \boldsymbol{V}_{ui}^T \Delta \boldsymbol{D} \boldsymbol{V}_{ui}$. The entries at the $i$-th row and $j$-th column of $\boldsymbol{F}_1$ and $\boldsymbol{F}_2$ are calculated as
\begin{equation}
    \begin{aligned}
        &\left[ \boldsymbol{F}_1 \right]_{i,j} = \sum_{m=1}^{N_f} {T_{ui}^{(m)}}^{i-1} {T_{ui}^{(m)}}^{j-1} \\
        &\left[ \boldsymbol{F}_2 \right]_{i,j} = \frac{1}{N_f} \left[\sum_{m=1}^{N_f} {T_{ui}^{(m)}}^{i-1} (-1)^{m}  \right] \left[\sum_{m=1}^{N_f} {T_{ui}^{(m)}}^{j-1} (-1)^{m-1}  \right]
    \end{aligned}
\label{eq59}
\end{equation}
We can observe that $\left[ \boldsymbol{F}_1 \right]_{i,j} \gg \left[ \boldsymbol{F}_2 \right]_{i,j}$, and this holds more strictly as $N_f$ increases. If $\boldsymbol{F}_2$ can be neglected compared to $\boldsymbol{F}_1$, then \eqref{eq58} reduces to 
\begin{equation}
    \boldsymbol{\Sigma}_{\boldsymbol{\gamma}_{u;ij}} = \sigma_n^2 \left(\boldsymbol{V}_{ui}^T \boldsymbol{V}_{ui} \right)^{-1} 
\label{eq60}
\end{equation}
Consequently, considering the TDOA estimates at the instants when target $u$ receives signals from anchor $i$, i.e., $\boldsymbol{T}_u= \boldsymbol{T}_{ui}$, and $\boldsymbol{V}_u = \boldsymbol{V}_{ui}$, the covariance matrix of the vector of TDOA estimates, namely the theoretical MSE, is derived by
\begin{equation}
\begin{split}
    \boldsymbol{\Sigma}_{\boldsymbol{\tau}_{u;ij}} &= \boldsymbol{V}_{ui} \boldsymbol{\Sigma}_{\boldsymbol{\gamma}_{u;ij}} \boldsymbol{V}_{ui}^T \\
    &= \sigma_n^2 \boldsymbol{V}_{ui} \left(\boldsymbol{V}_{ui}^T \boldsymbol{V}_{ui} \right)^{-1} \boldsymbol{V}_{ui}^T \\
    & = \text{CRLB2}(\boldsymbol{\tau}_{u;ij})
\end{split}
\label{eq61}
\end{equation}

To summarize, the conditions for the theoretical MSE of TDOA estimation to reach the CRLB are listed as follows\footnote{It is worth mentioning that our method can still achieve satisfactory performance without these conditions, as demonstrated by the simulation results in Section \textrm{VI}. These conditions are specifically raised for simplifying the analysis.}:
\begin{enumerate}[label=(\alph*)]
    \item $T_f \gg \left| i-j \right| T_s$, which indicates that the target's reception interval of signals from two anchors within the same frame is short enough compared to the duration of one frame.     \label{cond1}
    \item $\left[ \boldsymbol{F}_1 \right]_{i,j} \gg \left[ \boldsymbol{F}_2 \right]_{i,j}$, which is generally satisfied if $N_f$ is sufficiently large.        \label{cond2}
\end{enumerate}


We can interpret the two conditions as follows. Since concurrent measurements are not available in a TDBPS, we can only use sequential measurements to approximate the results. If $T_f \gg \left| i-j \right| T_s$, meaning that the target's reception interval of sequential measurements within the same frame is extremely short compared to the duration of one frame, the signals from anchor $i$ and anchor $j$ can be viewed as received almost simultaneously, approximating the scenario of concurrent measurements. Moreover, estimating the parameters of polynomial models resembles a fitting process, which requires data from more frames for greater accuracy.

Additionally, we observe an interesting case: if $L=1$ and $N_f$ is even, $\boldsymbol{F}_2$ is a scalar, which satisfies
\begin{equation}
    \boldsymbol{F}_2 =  \frac{1}{N_f} \left[\sum_{m=1}^{N_f} (-1)^{m}  \right] \left[\sum_{m=1}^{N_f} (-1)^{m-1}  \right] = 0
\label{eq62}
\end{equation}
This implies that the CRLB can be achieved as long as condition \ref{cond1} is satisfied under these circumstances.

\section{Numerical Simulation}

In this section, we conduct comprehensive numerical simulations to validate the effectiveness of our proposed method for TDOA estimation and its application to mobile target localization. We consider a 2D simulation scenario ($K=2$) involving a TDBPS composed of $N_a$ stationary anchors with known positions and well-synchronized clocks, and one mobile target $u$ with unknown position and asynchronous clock. Anchors broadcast signals following the measurement protocol introduced in Section \ref{problemFormulation}. Target $u$'s clock drift $\omega_u$ and initial clock offsets $\phi_u$ are random variables drawn from uniform distributions, specifically, $\omega_u \sim 1+\mathcal{U}(-20,20)$ parts per million (ppm) \cite{zhang2021signal} and $\phi_u \sim \mathcal{U}(-1,1)$ ms. The frame length $T_f$ is set to $0.1\text{ s}$ and each frame is divided into $N_s=20$ time slots whose duration is $T_s=5 \text{ ms}$. 
Unless otherwise specified, we assume that target $u$ is initially located at $(0,0)$\footnote{Coordinates are in meters, and the same applies below.}, and that $N_a=12$ anchors are all randomly placed within a square centered at $(0,0)$ with side lengths of 2 km.
Target $u$ moves with a constant velocity in a random direction. The speed is drawn from a uniform distribution, i.e., $v \sim \mathcal{U}(0,v_\text{max})$($v_\text{max}=10 \text{m/s}$). The transmission times are set by the anchors themselves, and their noise is generally only caused by clock jitter and can be neglected, which is a typical case in UWB systems \cite{xu2008delay,shi2020blas}. Therefore, we set $\sigma_t = 0$, while the variance of the noise in reception time measurements is set as $\sigma_r^2=10^{-3} \text{ m}^2$ \cite{xu2008delay}. Other parameters are specified in respective experiments.

\subsection{TDOA Estimation}\label{simTDOA}

We select two classical TDOA estimation methods as benchmarks for comparison, which represent the two previously mentioned approaches for compensating the impacts of target $u$'s clock drift:
\begin{itemize}
    \item \textbf{Benchmark 1 }(Asynchronous Time Difference of Arrival (A-TDOA)
    \cite{puatru2023flextdoa,dotlic2018ranging}) : A classical TDOA estimation method which requires CFO measurement assistance. Specifically, \cite{dotlic2018ranging} discusses the uplink TDOA case where targets transmit blink messages, while \cite{puatru2023flextdoa} extends the method to the downlink TDOA case, which is the focus of our problem. 
    \item \textbf{Benchmark 2 }(Signal-Multiplexing Network Ranging (SM-NR) \cite{zhang2021signal}): An effective network ranging method which is capable of both active ranging and differential ranging. 
\end{itemize}
Additionally, since the new P-TDOA utilizes information from multiple frames for estimation, whereas the benchmarks use only a single frame, we apply polynomial fitting to the raw TDOA estimates of the benchmarks for a fair comparison. This allows TDOAs to be estimated using a uniform polynomial model as in \eqref{eq17}. We refer to the fitting versions of the benchmarks as `A-TDOA*' and `SM-NR*', respectively. For the settings of each method, unless otherwise specified, the model order $L$ in P-TDOA is set to 2 and the solution period is set to $N_f =3$ frames; anchor 1 is designated as the `sync node' for SM-NR \cite{zhang2021signal}; and the CFO in A-TDOA is assumed to be measured with perfect accuracy. Moreover, the fitting order for `A-TDOA*' and `SM-NR*' is set equal to $L$. 

The root mean square error (RMSE) of the TDOA is given by
\begin{equation}
    \text{RMSE}_{\text{TDOA}} = \mathbb{E} \left \{ 
    \sqrt{ \frac{1}{N_f} {   \left\lVert \Delta \boldsymbol{\tau}_{u;i,j}   \right\lVert  }^2    }
    \right\}  
    \label{eq63}
\end{equation}
where $\Delta \boldsymbol{\tau}_{u;i,j}= \hat{\boldsymbol{\tau}}_{u;ij}- \boldsymbol{\tau}_{u;ij}$ represents the error vector of TDOA estimates between target $u$ and anchor pair $(i,j)$ across $N_f$ frames. Without loss of generality, we select the TDOAs between target $u$ and anchor pair $(1,2)(i=1,j=2)$ at the instants when target $u$ receives signals from anchor 1 for analysis, and conduct $N_{\text{sim}}=10000$ Monte-Carlo simulations for each type of experiments. In each Monte Carlo simulation, the anchor distribution, target dynamics, and clock parameters vary.
\begin{figure}[!tb]
    \centering
    \includegraphics[width=0.45\textwidth]{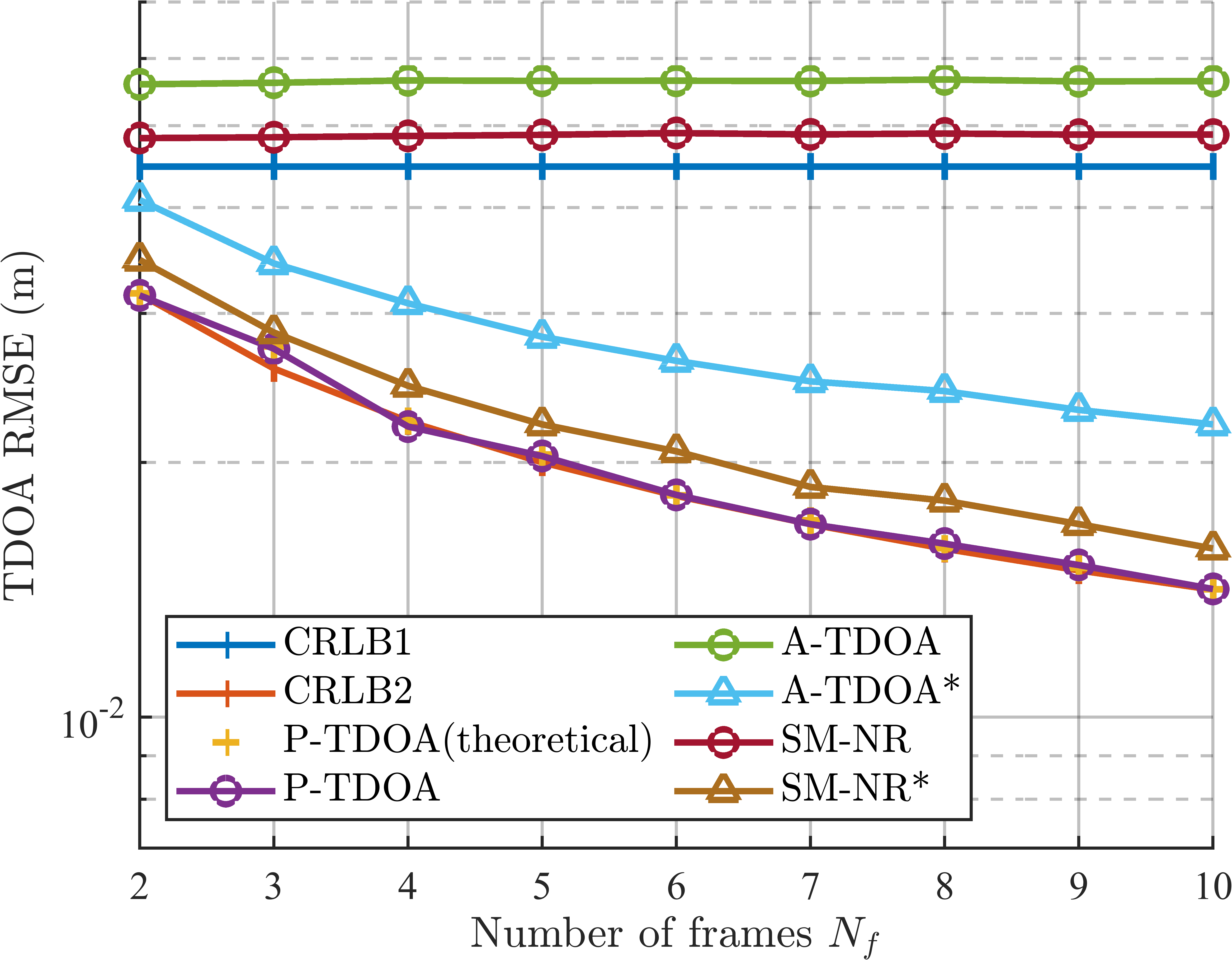}
    \caption{RMSE of TDOA estimates for stationary targets with varying numbers of frames.
}
\label{fig5}
\end{figure}

\subsubsection{Adaptability to Stationary Targets} \quad \\
\indent Although the P-TDOA method is designed for mobile targets, it can also be applied to stationary targets. In this experiment, we keep target $u$ stationary at $(0,0)$ and set $L=1$ for P-TDOA. The results are shown in Fig. \ref{fig5}. It is observed that the polynomial model reduces the CRLB for TDOA estimation ($\text{CRLB2}(\tau_{u;ij})<\text{CRLB1}(\tau_{u;ij})$), with $\text{CRLB2}(\tau_{u;ij})$ decreasing as the number of frames increases. Moreover, P-TDOA outperforms SM-NR, A-TDOA, and their respective fitting versions, surpassing $\text{CRLB1}(\tau_{u;ij})$ and asymptotically approaching $\text{CRLB2}(\tau_{u;ij})$.

Furthermore, it is noted that when $N_f$ is odd, e.g., $N_f = 3$, the theoretical RMSE of P-TDOA exhibits a slight deviation from $\text{CRLB2}(\tau_{u;ij})$. In contrast, when $N_f$ is even, the values align very closely, supporting the assertion behind \eqref{eq62}. The deviation decreases as $N_f$ increases  (e.g., $N_f = 5,7,9$), validating the previously stated condition \ref{cond2} in Section \ref{theoreticalMSE} for the theoretical MSE to reach $\text{CRLB2}(\tau_{u;ij})$.

\begin{figure}[!tb]
    \centering
    \subfloat[RMSE of TDOA estimates for moving targets with varying reception intervals.]{
        \includegraphics[width=0.45\textwidth]{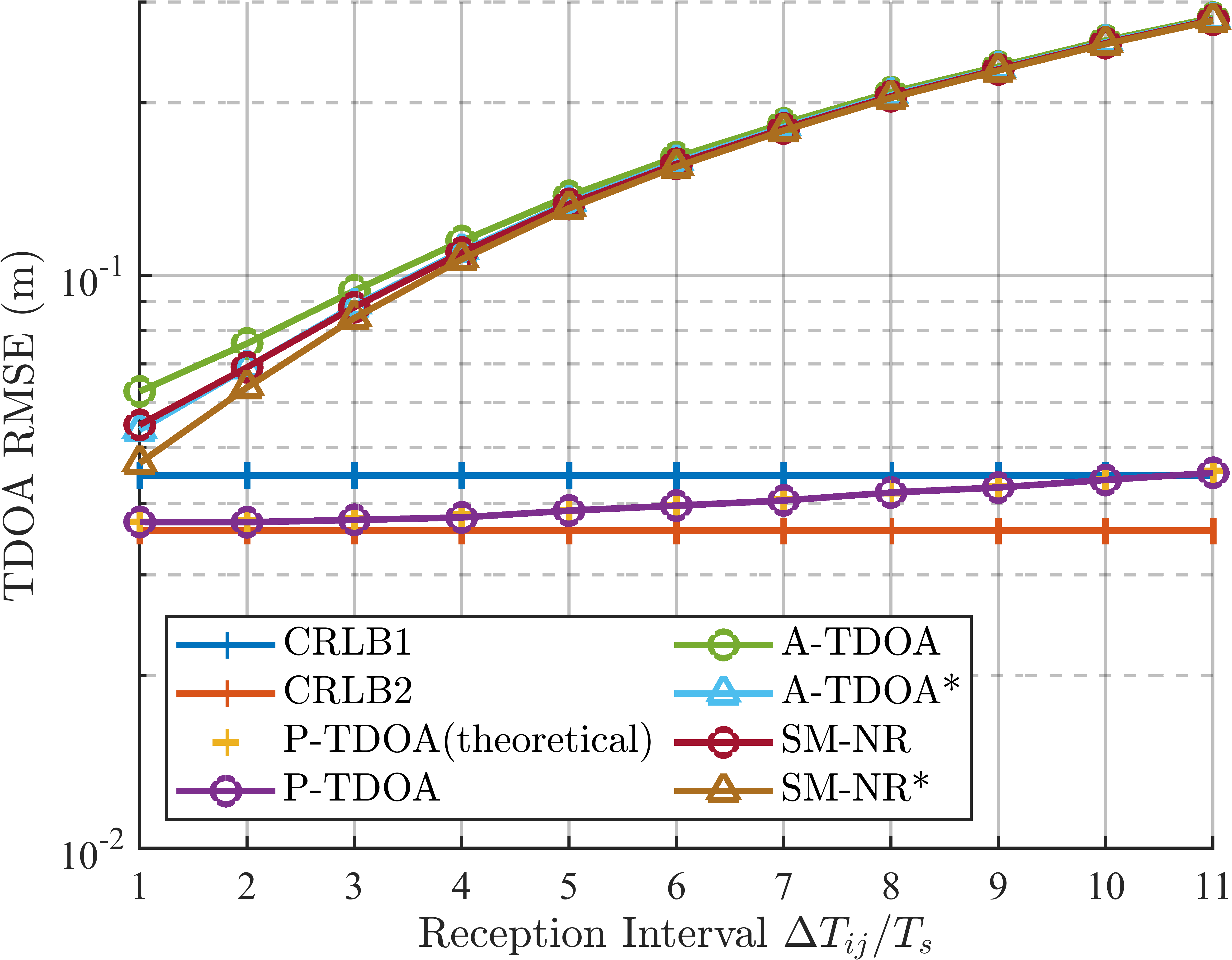}
        \label{fig6_a}
    }\\
    \subfloat[Boxplots of TDOA estimation errors for moving targets with varying reception intervals.]{
        \includegraphics[width=0.45\textwidth]{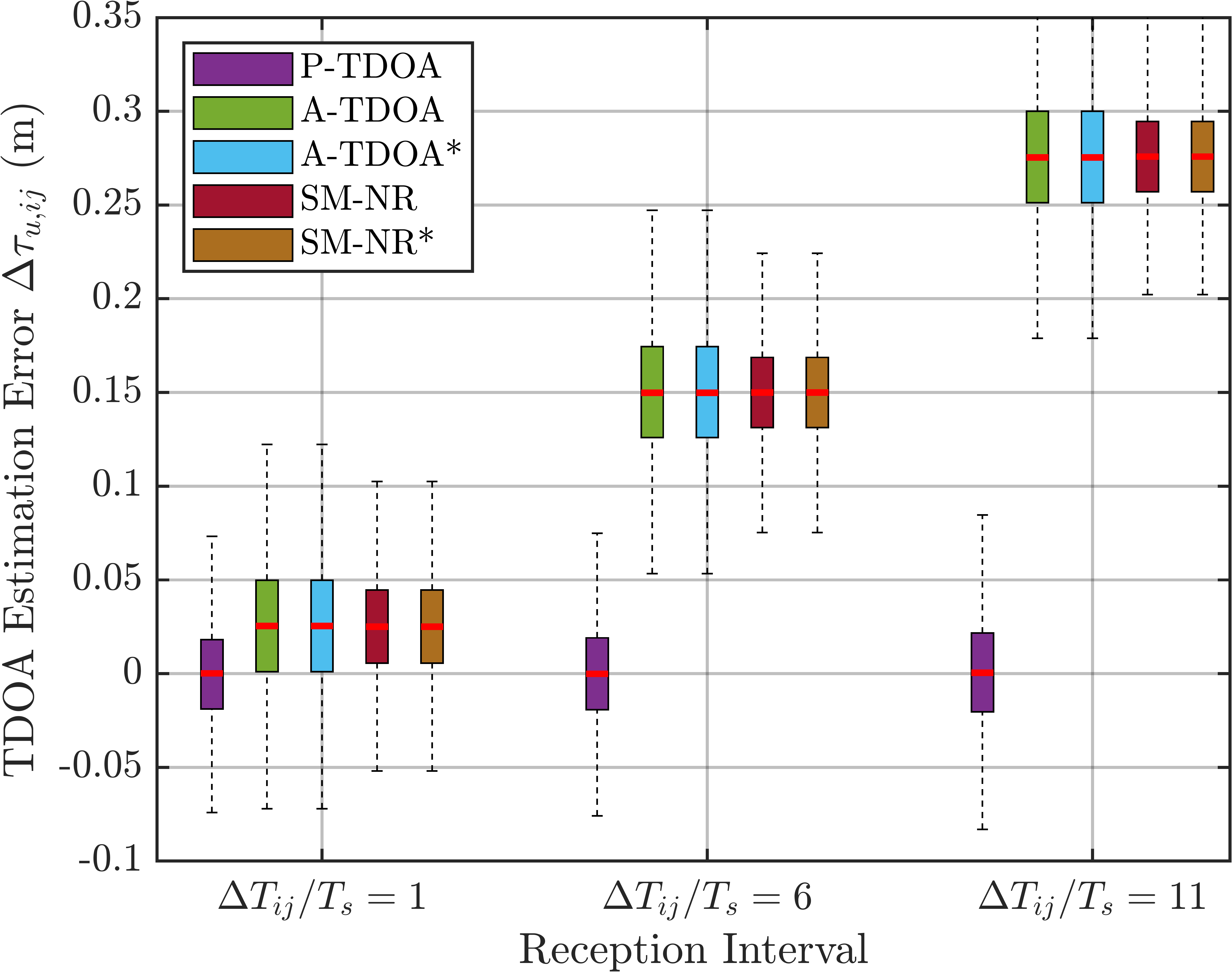}
        \label{fig6_b}
    }
    \caption{Comparison of TDOA estimates for moving targets with varying reception intervals.}
    \label{fig6}
\end{figure}

\begin{figure*}[!tb]
\centering
\subfloat[\centering Motion 1]{\includegraphics[width=0.3\textwidth]{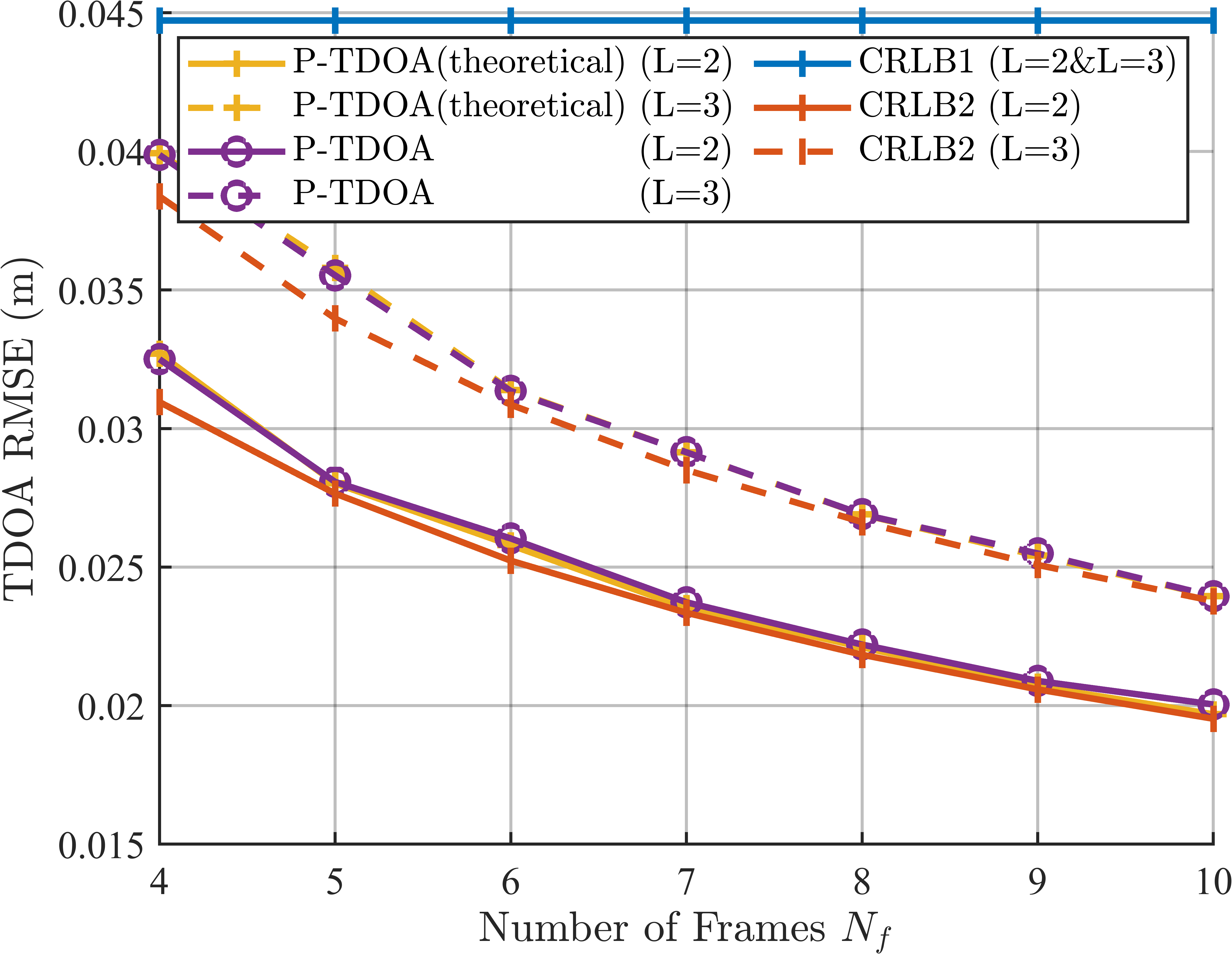}%
\label{fig7_a}}
\hfil
\subfloat[\centering Motion 2]{\includegraphics[width=0.3\textwidth]{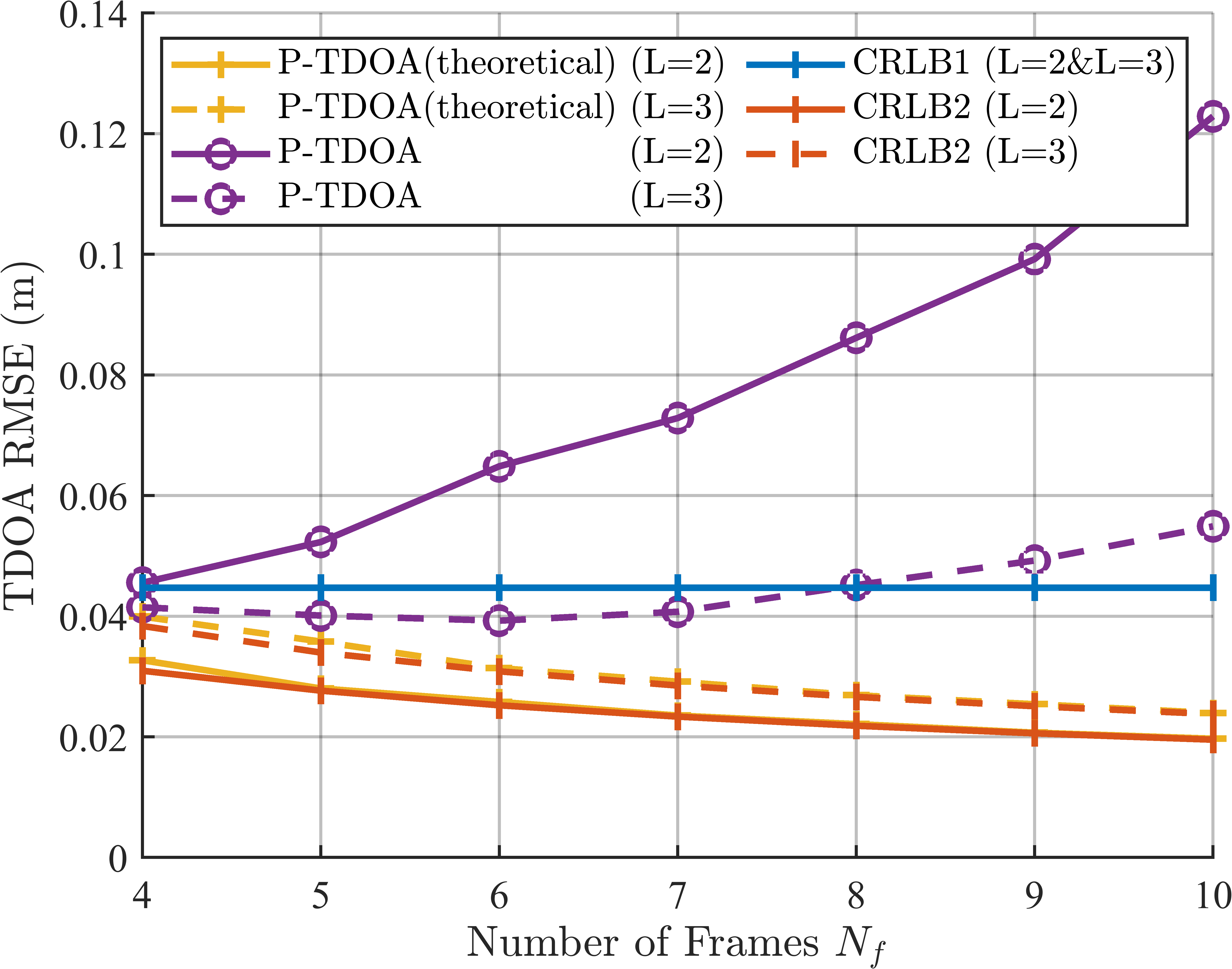}%
\label{fig7_b}}
\hfil
\subfloat[\centering Motion 3]{\includegraphics[width=0.3\textwidth]{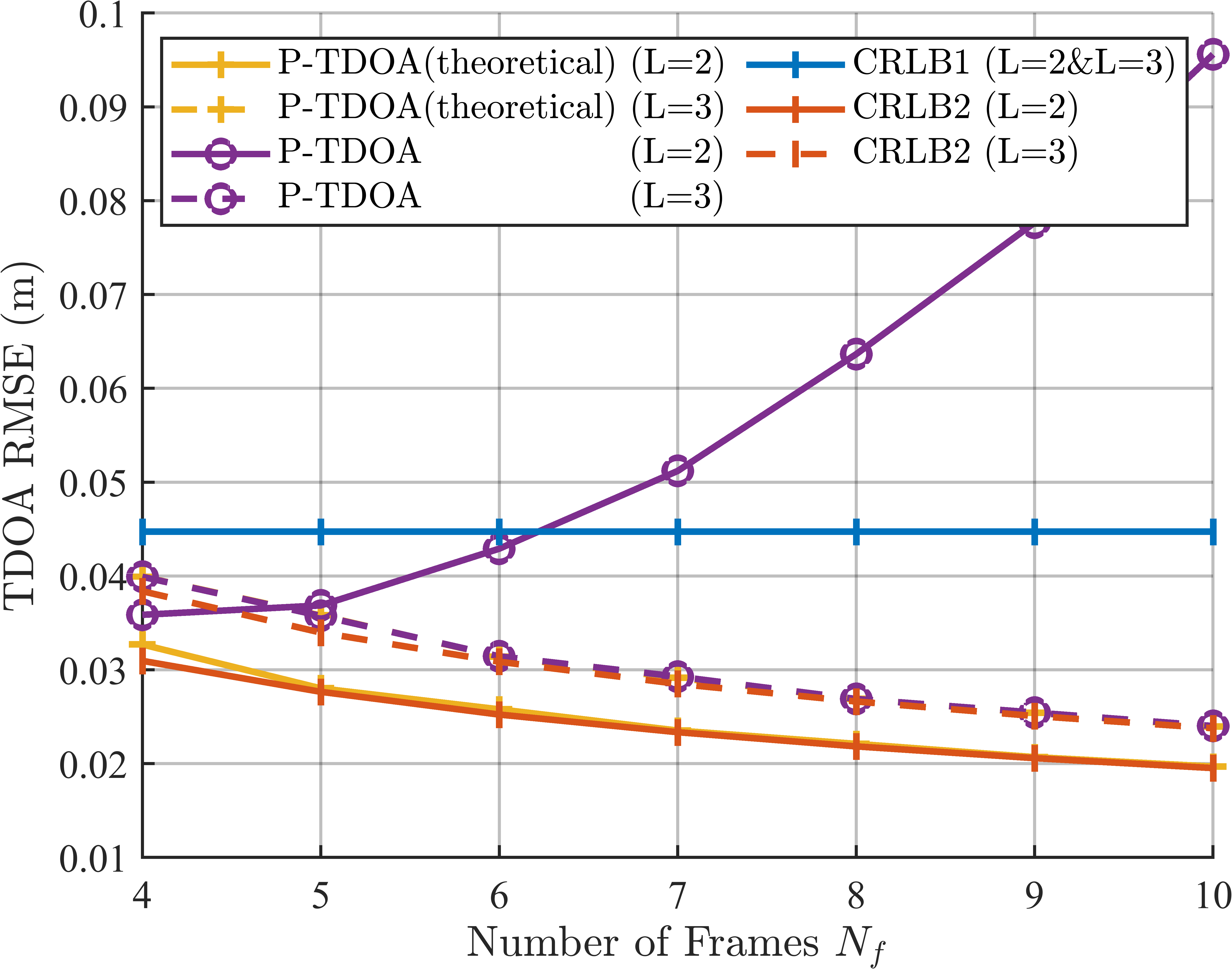}%
\label{fig7_c}
}
\caption{RMSE of TDOA estimates using P-TDOA for moving targets with varying motion types (a) Motion 1: uniform linear motion, (b) Motion 2: uniform circular motion, (c) Motion 3: uniformly accelerated linear motion. 
}
\label{fig7}
\end{figure*}

\subsubsection{Adaptability to Varying Reception Intervals}  \quad \\
\indent 
From this part on, we focus on evaluating the performance of P-TDOA for mobile targets. As previously mentioned, for a mobile target, the displacement during the reception interval causes the variation of distances between the target and anchors, making the transmission and reception times recorded at different moments not correspond to the TDOA at any specific instant. If the target moves with a constant velocity, varying reception intervals lead to varying displacements of the target during those intervals, potentially affecting the performance for TDOA estimation. Thus, we use this experiment to test the estimation error against varying reception intervals.

However, in our protocol, the reception intervals between signals from anchor 1 and anchor 2 remain nearly constant. Therefore, we assume that all anchors, except for anchor 1, are placed in the same position. In this way, measuring the TDOA between target $u$ and anchor pairs of anchor 1 with other anchors effectively represents the scenarios of measuring the TDOA between target $u$ and anchor pair $(1,2)$ with varying signal reception intervals.

We place anchor 1 and anchor $j$($j=2,3,\ldots,12$) at locations $(1000,0)$ and $(0,1000)$, respectively. The target starts from $(0,0)$ and moves toward anchor $j$ at a constant speed of $v=5\text{ m/s}$. The RMSE of the TDOA estimates with varying reception intervals is shown in Fig. \ref{fig6}(a). It can be observed that for A-TDOA, SM-NR, and their respective fitting versions, the RMSE of TDOA estimates increases significantly as the reception interval increases. In contrast, although P-TDOA shows a similar trend, the error grows considerably slower. Moreover, with smaller reception intervals, P-TDOA approaches the $\text{CRLB2}(\tau_{u;ij})$, confirming the condition \ref{cond1} given in Section \ref{theoreticalMSE}.

Additionally, we analyze the distribution of TDOA estimation errors for each method and plot boxplots for three selected reception intervals, as shown in Fig. \ref{fig6}(b). From the red lines which indicate the mean error\footnote{In a typical boxplot, the line within the box usually represents the median. Here, we have substituted the median with the mean to highlight whether there is any bias in the estimates.}, it can be seen that P-TDOA produces unbiased TDOA estimates, whereas estimates from other methods exhibit bias. 
Furthermore, this bias grows with increasing reception intervals (i.e., greater displacements of the target). It demonstrates that traditional methods become ineffective when the target moves at high speeds or when there are larger reception intervals between anchors, whereas P-TDOA remains applicable in these scenarios.

\subsubsection{Adaptability to Different Motion}  \quad \\ 
\indent Previous experiments have demonstrated that P-TDOA outperforms the benchmarks in TDOA estimation for mobile targets. In this part, we evaluate the adaptability of P-TDOA to different motion. We consider three types of motion:
\begin{enumerate}[label=\textcircled{\arabic*}]
    \item Motion 1 (uniform linear motion):
    Target $u$ moves with a constant velocity and the speed is drawn from a uniform distribution, i.e.,
    $v\sim \mathcal{U}(0, v_{\text{max}})$($v_{\text{max}} = 10\text{ m/s}$), which is consistent with the default motion.
    \item Motion 2 (uniform circular motion): Target $u$ moves in a circle with constant speed drawn from a uniform distribution, i.e.,
    $v\sim \mathcal{U}(0, v_{\text{max}})$($v_{\text{max}} = 10\text{ m/s}$). The circle radius $R$ is also drawn from a uniform distribution with $R\sim \mathcal{U}(0, R_\text{max})$($R_\text{max}=100$ m). 
    \item Motion 3 (uniformly accelerated linear motion): Target $u$ moves in a line with constant acceleration. The initial speed is drawn from a uniform distribution, i.e.,
    $v\sim \mathcal{U}(0, v_{\text{max}})$($v_{\text{max}} = 10\text{ m/s}$), and the speed acceleration $a \sim \mathcal{U}(0, a_\text{max})$($a_\text{max}=5 \text{ m/s}^2$).
\end{enumerate}

For each type of motion, we apply $L=2$ and $L=3$ in P-TDOA for comparison. As illustrated in Fig. \ref{fig7}, we observe that the theoretical RMSE for lower-order $L = 2$ is smaller than those for higher-order $L = 3$ across all three types of motion. If the model accurately captures the variation of the TDOA, the estimation accuracy improves with an increasing number of frames, as seen in the $L = 2$ and $L = 3$ cases for Motion 1 and the $L = 3$ case for Motion 3. However, if the model fails to match the variation of the TDOA, the actual RMSE deviates from the theoretical value, and an increasing number of frames may degrade estimation performance, as demonstrated by the $L = 2$ and $L = 3$ cases in Motion 2 and the $L = 2$ case in Motion 3. Therefore, in practical implementation, careful selection of the model order and the number of frames for estimation is crucial. A feasible approach is order recursive least squares (ORLS) as discussed in \cite{rajan2015joint,kay1993fundamentals}, which is beyond the scope of this paper and will not be elaborated further.

\subsubsection{Adaptability to Varying Noise Levels} 
\begin{figure}[!tb]
\centering
{\includegraphics[width=0.45\textwidth]{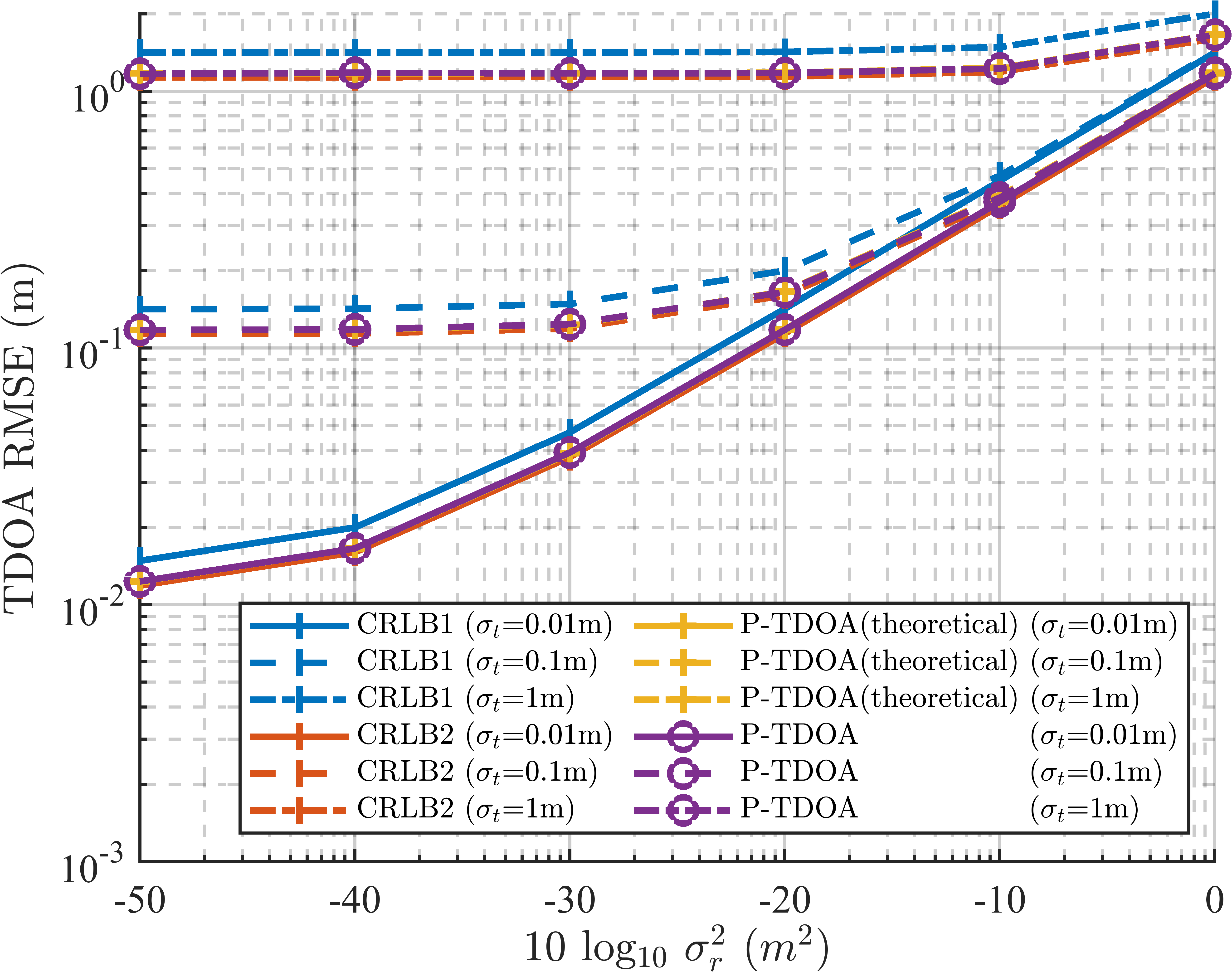}}
\caption{RMSE of TDOA estimates using P-TDOA for moving targets under different noise levels.
}
\label{fig8}
\end{figure}

\quad \\
\indent In this part, we examine the impact of varying noise levels in time measurements on TDOA estimation. As shown in Fig. \ref{fig8}, the results confirm the accuracy of our theoretical analysis and demonstrate the effectiveness of the proposed P-TDOA method across varying noise levels in transmission and reception time measurements, with estimation accuracy surpassing $\text{CRLB1}(\tau_{u;ij})$ and approaching $\text{CRLB2}(\tau_{u;ij})$.

\subsection{Mobile Target Localization} \label{simLoc}

With P-TDOA, concurrent TDOA estimates can be derived by assigning the same instant to the estimated TDOA models of multiple anchor pairs. The target's position is then determined by solving a set of hyperbolic equations, each representing a hyperbolic curve with the foci at the corresponding anchor pairs. Numerous algorithms have been developed to solve this problem \cite{chan1994simple,sun2011asymptotically,ghany2019parametric,kravets2024new}. Chan \textit{et al.} provide a closed-form solution that approximates the ML estimator under low noise conditions, requiring a minimum of three anchors for localization in 2D scenarios \cite{chan1994simple}. Sun \textit{et al.} extend this approach to account for uncertainties in anchor positions \cite{sun2011asymptotically}. Additionally, challenging scenarios involving measurement outliers and unfavorable anchor distributions are addressed in \cite{ghany2019parametric} and \cite{kravets2024new}, respectively.

In this subsection, we employ the algorithm in \cite{chan1994simple} to demonstrate the effectiveness of our method for mobile target localization. Other TDOA-based localization techniques can also be directly applied to different scenarios. For detailed information on the multilateration algorithm, readers can refer to \cite{chan1994simple}.

We select two recent sequential measurement-based mobile target localization methods as benchmarks for comparison, which have similar problem settings and application scopes to our proposed method.
\begin{itemize}
    \item \textbf{Benchmark 1} (Closed-Form Joint Localization and Synchronization (CFJLAS)\cite{guo2022new}): A closed-form mobile target localization method that uses sequential measurements from one-way broadcast signals. It assumes that the mobile target moves at a constant velocity during a frame, and jointly estimates position, velocity, and clock parameters.
     \item \textbf{Benchmark 2} (Timestamp-Conversion-based Joint Localization and Synchronization (TCJLAS) \cite{sun2024novel}): A method that utilizes timestamp conversion, employing Lagrange interpolation to transform sequential measurements into concurrent measurements for joint localization and synchronization.
\end{itemize}

Since the localization method in \cite{chan1994simple} employs a two-step WLS method, involving a series of linear operations. The final localization error $\Delta \boldsymbol{p}_u$ is related to the TDOA estimation error $\Delta \tau_u$ through a linear mapping, represented by the matrix $\boldsymbol{G}$, i,e., $\Delta p_u = \boldsymbol{G} \Delta \tau_u$. Thus, since the TDOA estimates provided by P-TDOA are unbiased and zero-mean, the final position estimates are also unbiased. Therefore, we also provide the CRLB for localization as a baseline for comparison, which is derived in \cite{chan1994simple}. It's worth mentioning that the CRLB for localization depends on the covariance matrix of TDOA estimates. Given that we impose polynomial model constraints on the TDOA estimates, these constraints must also be accounted for in the CRLB for localization. Similar to the CRLBs for TDOA estimation, we define the CRLB for localization without polynomial model constraints as $\text{CRLB1}(\boldsymbol{p}_u)$, and the CRLB considering these constraints as $\text{CRLB2}(\boldsymbol{p}_u)$. The simple deviation of the CRLBs for localization is given in Appendix \ref{appendix2}.

The default system settings are the same as those in Section \ref{simTDOA}. Since the anchors are placed randomly, the solution may degrade for certain anchor distributions. To better evaluate the performance of different methods, we exclude outliers based on unified thresholds. The RMSE of the position is given by
\begin{equation}
    \text{RMSE}_{\text{p}} = \mathbb{E} \left \{ 
    \sqrt{ \frac{1}{N_f}  \sum_{m=1}^{N_f} {   \left\lVert \hat{\boldsymbol{p}}_u^{(m)} -  \boldsymbol{p}_u^{(m)}  \right\lVert  }^2    }
    \right\}  
    \label{eq64}
\end{equation}
where $\hat{\boldsymbol{p}}_u^{(m)}$ and $\boldsymbol{p}_u^{(m)}$ represent the estimated position of target $u$ and its ground truth at the beginning of frame $m$. For each type of experiment, we conduct $N_{\text{sim}}=10000$ Monte-Carlo simulations. In each Monte Carlo simulation, the anchor distribution, target dynamics, and clock parameters vary.

\subsubsection{Performance with Varying Noise Levels} 
\begin{figure}
    \centering    \includegraphics[width=0.45\textwidth]{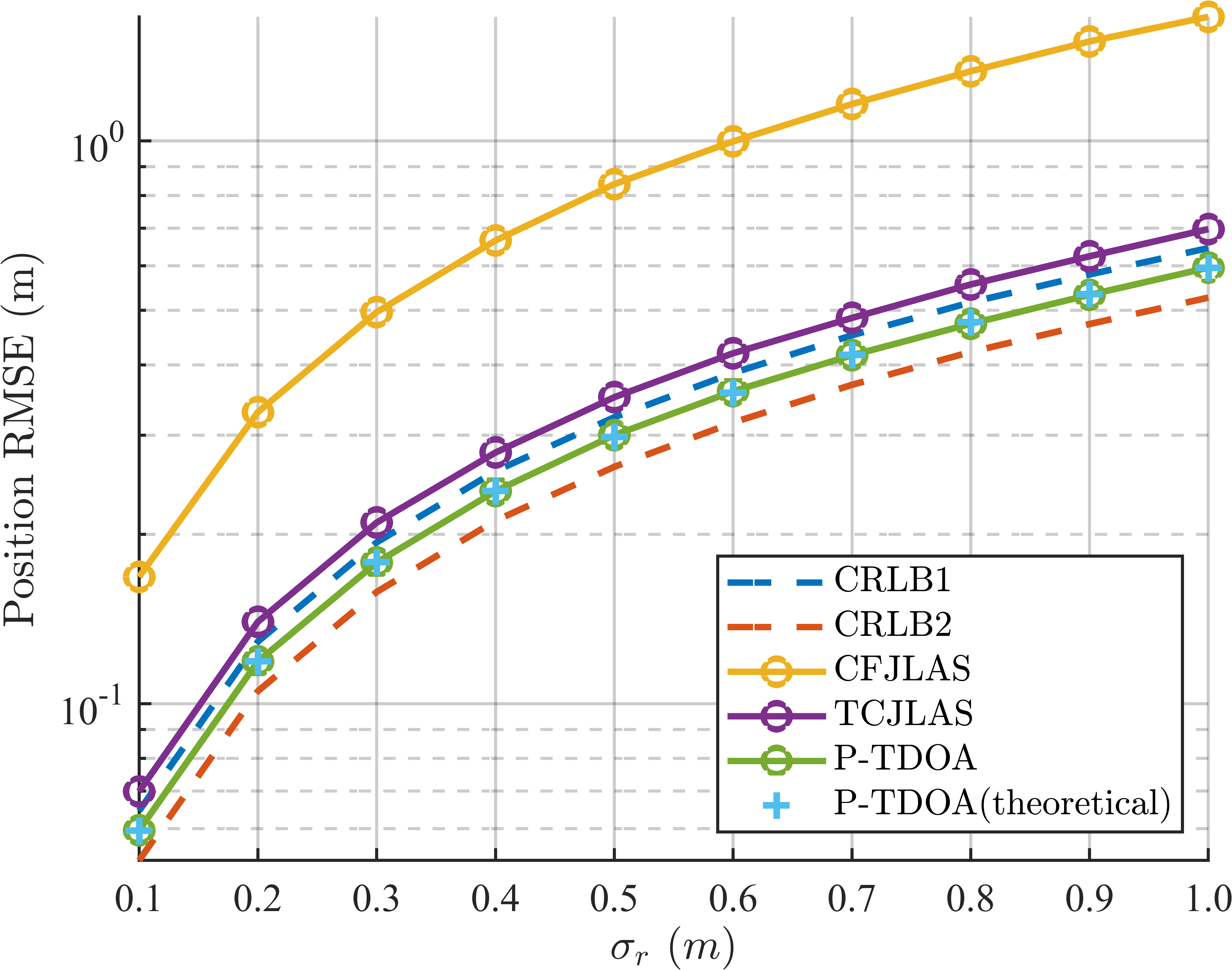}
    \caption{RMSE of position estimates for moving targets with varying noise levels.}
    \label{fig9}
\end{figure}

\quad \\
\indent We vary the reception time measurement noise from $0.1\text{ m}$ to $1\text{ m}$, with increments of $0.1\text{ m}$. As illustrated in Fig. \ref{fig9}, it is evident that the polynomial model used for TDOA estimation also reduces the CRLB for localization. The proposed P-TDOA method outperforms the benchmark methods, surpassing $\text{CRLB1}(\boldsymbol{p}_u)$ and approaching $\text{CRLB2}(\boldsymbol{p}_u)$ under low noise conditions, which is consistent with the characteristics of the method described in \cite{chan1994simple}.

\subsubsection{Performance with Varying Number of Anchors} 
\begin{figure}
    \centering
    \includegraphics[width=0.45\textwidth]{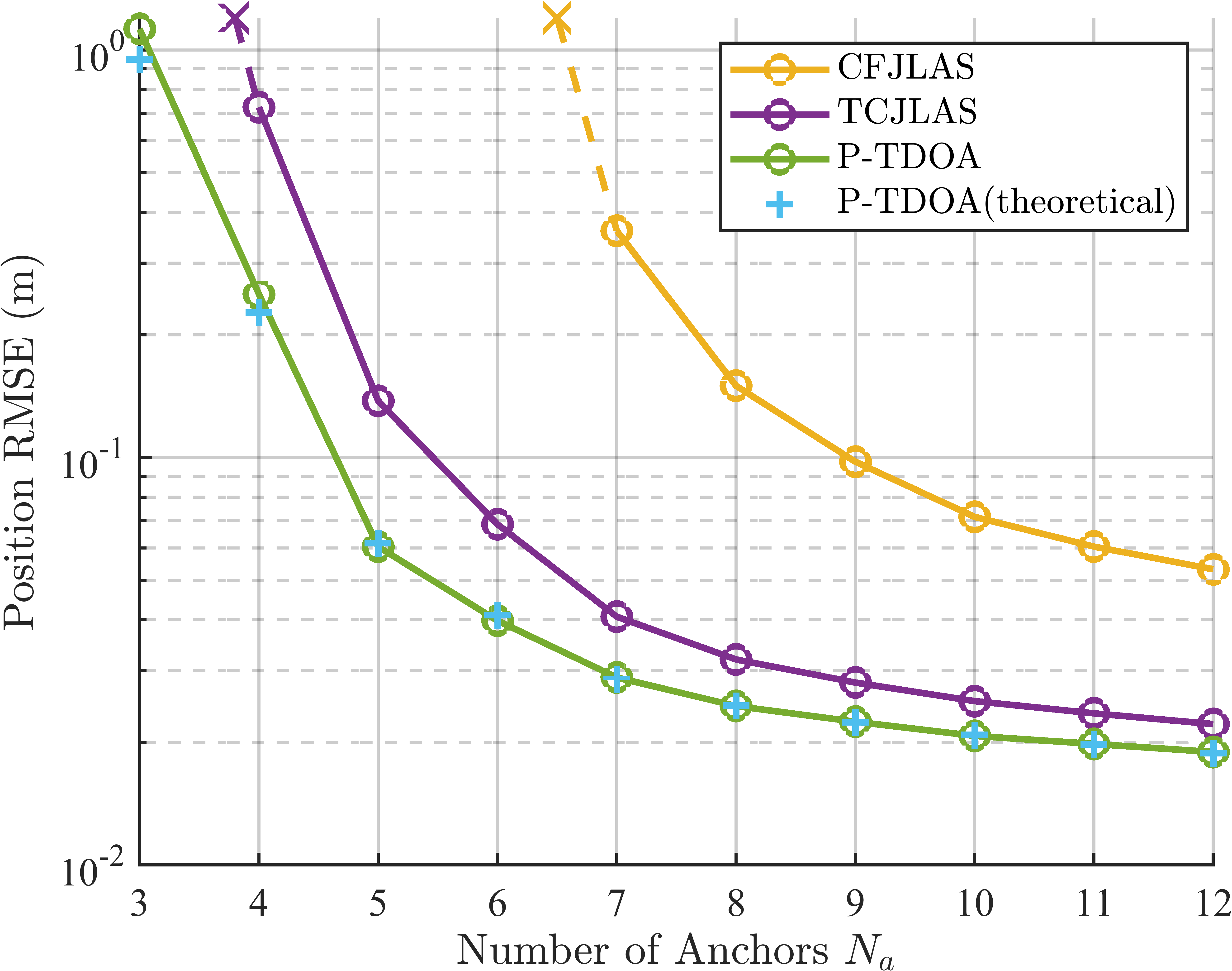}
    \caption{RMSE of position estimates for moving targets with varying number of anchors. The X marks indicate that the methods don't work with a smaller number of anchors. P-TDOA requires a minimum of 3 anchors for 2D localization, while TCJLAS and CFJLAS need at least 4 and 7 anchors, respectively.
    }
    \label{fig10}
\end{figure}

\quad \\
\indent The number of anchors plays a crucial role in determining localization performance. We vary the number of anchors from 3 to 12, and  present the corresponding position estimation RMSE in Fig.
\ref{fig10}. The proposed P-TDOA method requires a minimum of three anchors for 2D localization\footnote{The position RMSE of P-TDOA exhibits a slight deviation from the theoretical value when using only three or four anchors. This occurs because, with such a limited number of anchors, the solution provided by the algorithm in \cite{chan1994simple} becomes sensitive to the geometric distribution of the anchors. Despite filtering out some outliers, there remain data points with significant errors, which distort the average errors away from the theoretical values.}, while TCJLAS and CFJLAS need at least four and seven anchors, respectively, due to the joint estimation of additional parameters. With the same number of anchors, P-TDOA consistently outperforms the benchmark methods, and the localization accuracy improves as the number of anchors increases. Thus, P-TDOA not only reduces the number of required anchors but also achieves superior performance with a fixed number of anchors.

The reduced anchor requirement is attributed to the decoupling of localization from synchronization and target dynamics. Our proposed method leverages data in multiple frames to eliminate the unknown clock parameters of target $u$ in Section \ref{clockEliminate} and capture the dynamics of TDOAs in Section \ref{TDOAmodel}. Essentially, our approach exchanges a longer solution period for fewer anchor requirements. Since maintaining a stable topology over short periods is more feasible than establishing multiple connections, our method is well-suited for a wide range of application scenarios.

\subsubsection{Performance with Varying Anchor Distributions} 
\begin{figure}
    \centering
    \includegraphics[width=0.45\textwidth]{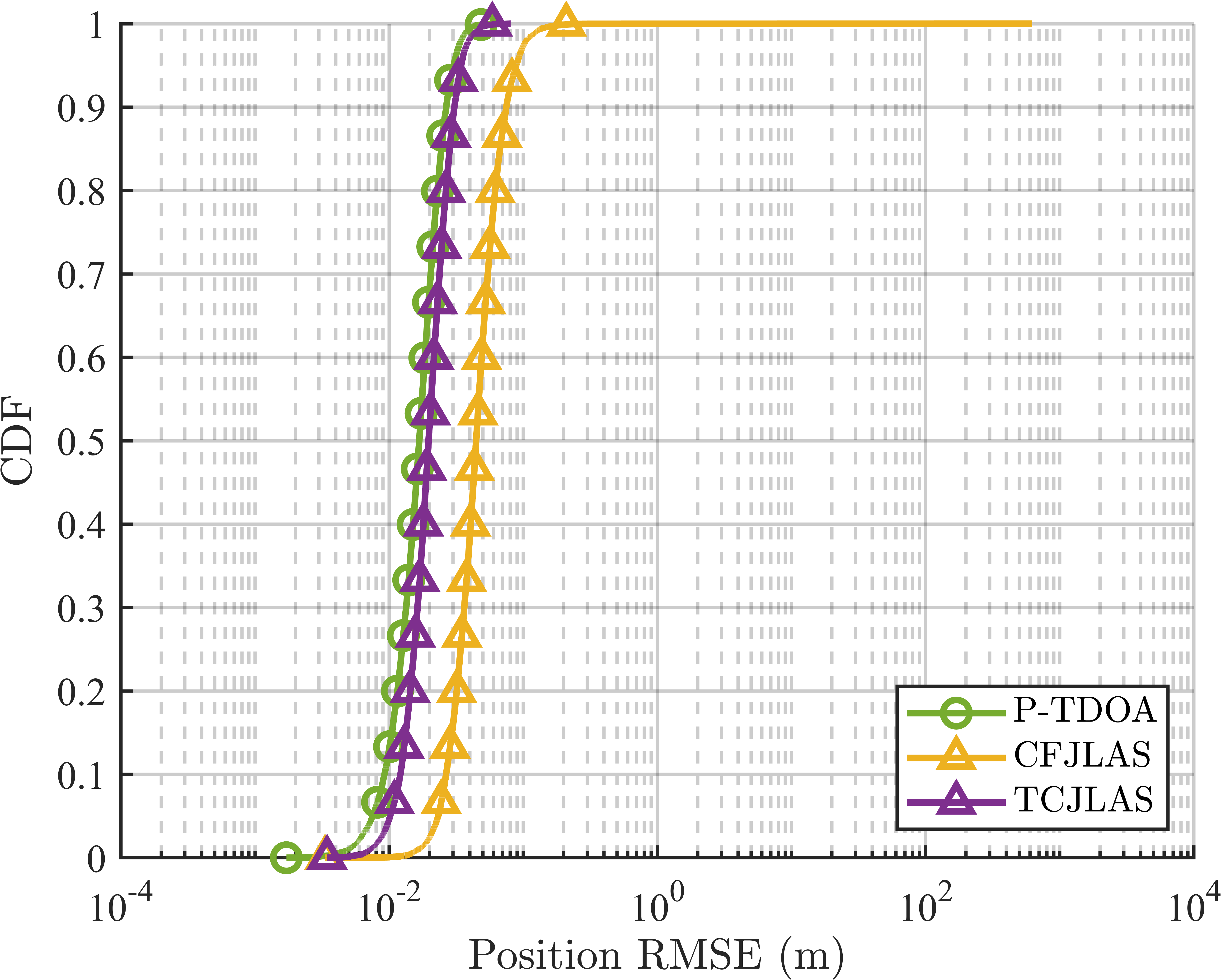}
    \caption{Cumulative distribution curves of RMSE for position estimates with different anchor point distributions.
    }
    \label{fig11}
\end{figure}

\quad \\
\indent The spatial distributions of anchors significantly affects localization performance. Fig. \ref{fig11} shows the cumulative distribution function (CDF) of the position MSE for different methods under random anchor distributions set in all our numerical simulations. Notably, the curve for P-TDOA is the steepest among all compared methods and achieves the lowest estimation MSE when the CDF reaches 1. These findings demonstrate the superior adaptability and robustness of P-TDOA to varying anchor distributions.

\subsubsection{Performance with High Target Dynamics} 
\begin{figure}
    \centering
    \includegraphics[width=0.45\textwidth]{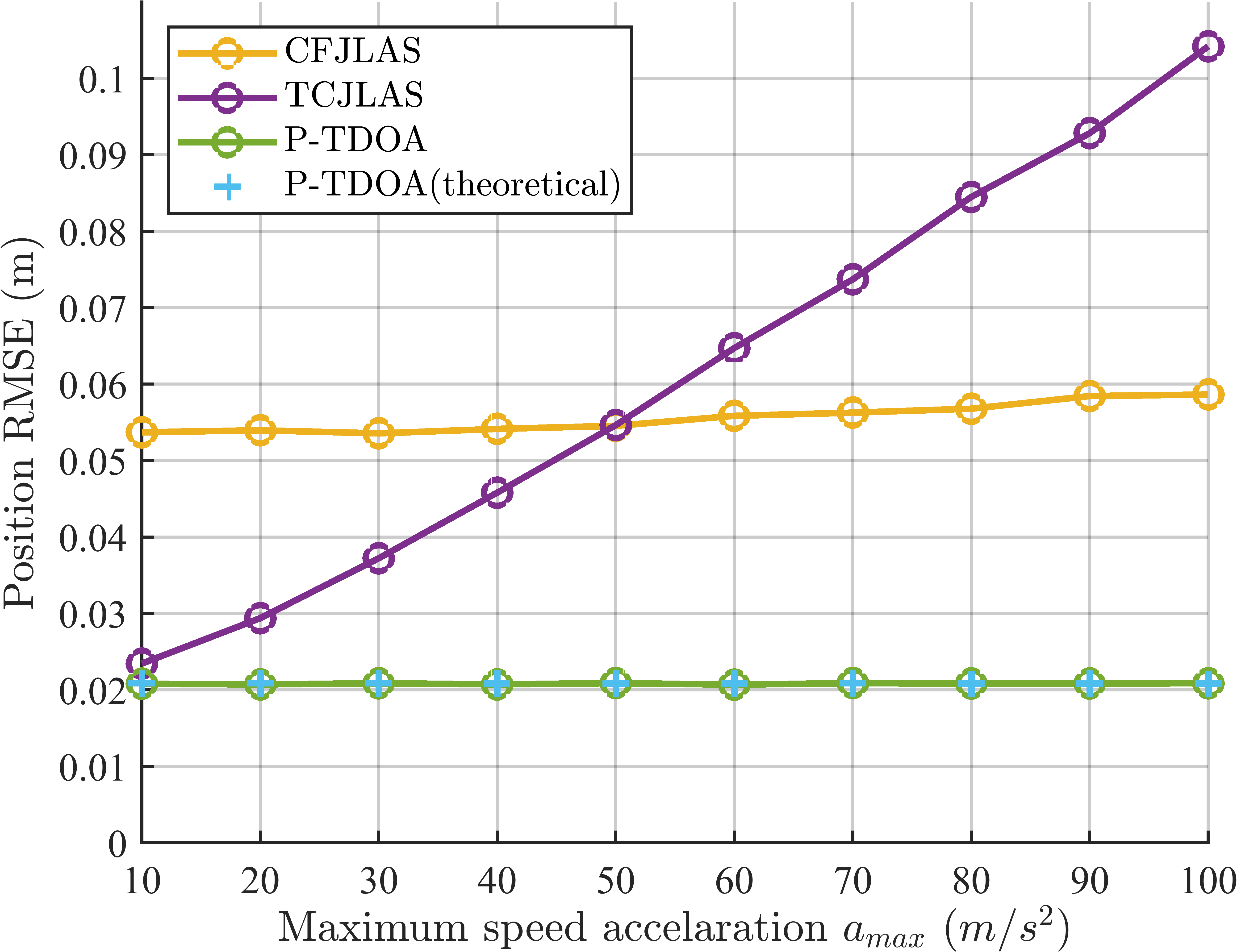}
    \caption{RMSE of position estimates with high target dynamics.
    }
    \label{fig12}
\end{figure}

\quad \\
\indent In the previous localization experiments, the target was set to move with a constant velocity. However, there are many scenarios where mobile targets exhibit high dynamics, such as during takeoff or landing. Here, we consider the target undergoing uniformly accelerated linear motion with high dynamics. The initial speed and acceleration are independently drawn from uniform distributions, i.e., $v \sim \mathcal{U}(0,v_\text{max})$ ($v_\text{max} = 10$ m/s) and $a \sim \mathcal{U}(0, a_\text{max})$, respectively. We set $a_\text{max}$ to vary from $10\text{ m/s}^2$ to $100\text{ m/s}^2$, with an increment of 10 $\text{m/s}^2$. For P-TDOA, we set $L = 3$ and $N_f = 4$. Fig. \ref{fig12} presents the position RMSE for varying accelerations. The results demonstrate that P-TDOA outperforms the benchmark methods in high-dynamic scenarios. Moreover, we observe that the performance of TCJLAS degrades significantly as acceleration increases. This degradation occurs because TCJLAS implicitly assumes a constant velocity within each frame to obtain concurrent measurements, making it unsuitable for highly dynamic situations. Although CFJLAS also assumes constant velocity, it does not directly alter the original measurements, and the duration for constant velocity assumption is restricted to $N_a T_s < N_s T_s = T_f$, which mitigates the impact of acceleration compared to TCJLAS.

\section{Conclusion}

In this paper, we propose a novel parameterized method for TDOA estimation of mobile targets in a TDBPS, namely P-TDOA. P-TDOA leverages data across multiple frames to eliminate unknown clock parameters and employs a polynomial model to approximate time-varying TDOAs. Through dedicated transformations, we transform the equations involving unknown propagation times into equations in terms of the unknown TDOA model parameters. We then introduce a WLS approach called MWLS to estimate the model parameters. To ensure the validity of the TDOA model parameter equations and a well-posed WLS problem, we propose an equation construction strategy named STDS. With this new method, TDOA estimates at arbitrary times within the solution period can be derived by assigning the desired instant to the TDOA model, enabling the derivation of concurrent TDOA estimates for all anchor pairs, and the subsequent application of classical TDOA localization methods. The new method builds a bridge between the TDBPS and all the classical TDOA-based localization methods.

We perform comprehensive theoretical analysis and extensive simulations, demonstrating that the proposed P-TDOA method outperforms existing methods and closely approaches the CRLB for TDOA estimation. Building upon the concurrent TDOA estimates provided by P-TDOA, the TDOA-based localization approach for mobile targets exhibits several advantages over state-of-the-art sequential measurement-based methods. First, it achieves higher localization accuracy. Second, it reduces the number of required anchors, with only three anchors being sufficient in 2D scenarios. Finally, the solution process involves only simple linear operations, avoiding complex nonlinear optimization or iterative calculations, which makes the method computationally efficient and easy to implement.

Several aspects warrant further investigation. First, since P-TDOA utilizes data from multiple frames, real-time performance is affected. We will explore strategies to enhance the adaptability and efficiency of P-TDOA in real-time applications. Additionally, while P-TDOA successfully eliminates unknown clock parameters and focuses on localization, clock synchronization for passive targets remains an open issue. Future research will focus on how to achieve clock synchronization within the P-TDOA framework.


\begin{appendices}
\section{Derivation of the specified $\boldsymbol{\Sigma}_{\boldsymbol{\eta}_{u;ij}}$} \label{appendix1}
We derive the specified $\boldsymbol{\Sigma}_{\boldsymbol{\eta}_{u;ij}}$ with the periodic broadcast protocol and STDS applied. 
The vectors of the noise in the transmission and reception times across $N_f$ frames are denoted as
\begin{equation}
\begin{aligned}
    \boldsymbol{v}_i = [v_{i}^{(1)},\ldots, v_{i}^{(N_f)}]^T
    &, \quad
    \boldsymbol{v}_j = [v_{j}^{(1)},\ldots, v_{j}^{(N_f)}]^T \\
    \boldsymbol{w}_{ui} = [w_{ui}^{(1)},\ldots, w_{ui}^{(N_f)}]^T
    &, \quad
    \boldsymbol{w}_{uj} = [w_{uj}^{(1)},\ldots, w_{uj}^{(N_f)}]^T
\end{aligned}
\label{eq65}
\end{equation}
The combined noise in \eqref{eq25} can then be specified by
\begin{equation}
\boldsymbol{\eta}_{u;ij} =  \boldsymbol{C}_{u;ij}^{\boldsymbol{v}_i} \boldsymbol{v}_i +  \boldsymbol{C}_{u;ij}^{\boldsymbol{v}_j} \boldsymbol{v}_j +
\boldsymbol{C}_{u;ij}^{\boldsymbol{w}_{ui}} \boldsymbol{w}_{ui}+ 
\boldsymbol{C}_{u;ij}^{\boldsymbol{w}_{uj}} \boldsymbol{w}_{uj}
\label{eq66}
\end{equation}
where the entries in $\boldsymbol{C}_{u;ij}^{\boldsymbol{v}_i}$,$\boldsymbol{C}_{u;ij}^{\boldsymbol{v}_j}$, $\boldsymbol{C}_{u;ij}^{\boldsymbol{w}_{ui}}$, $\boldsymbol{C}_{u;ij}^{\boldsymbol{w}_{uj}}$ are all zero except for the following items:
\begin{equation}
\begin{aligned}
     &\left[ \boldsymbol{C}_{u;ij}^{\boldsymbol{v}_i} \right]_{s,s:s+1} = \left[ -(\hat{T}_{ui}^{(s+1)} - \hat{T}_{uj}^{(s)}) ,  (\hat{T}_{ui}^{(s)} - \hat{T}_{uj}^{(s+1)}) \right] \\
     &\left[ \boldsymbol{C}_{u;ij}^{\boldsymbol{v}_j} \right]_{s,s:s+1} = \left[ -(\hat{T}_{ui}^{(s)} - \hat{T}_{uj}^{(s+1)}) ,  (\hat{T}_{ui}^{(s+1)} - \hat{T}_{uj}^{(s)}) \right] \\
    &\left[ \boldsymbol{C}_{u;ij}^{\boldsymbol{w}_{ui}} \right]_{s,s:s+1} = \left[ 
     (\hat{T}_i^{(s+1)}  - \hat{T}_j^{(s)}),  -(\hat{T}_i^{(s)} - \hat{T}_j^{(s+1)})   \right] \\
    &\left[ \boldsymbol{C}_{u;ij}^{\boldsymbol{w}_{uj}} \right]_{s,s:s+1} = \left[ 
     (\hat{T}_i^{(s)} - \hat{T}_j^{(s+1)}), - ( \hat{T}_i^{(s+1)} - \hat{T}_j^{(s)}) 
     \right]
\end{aligned}
\label{eq67}
\end{equation}
where $s=1,2,\ldots,N_f-1$. Thus, the covariance matrix of $\boldsymbol{\eta}_{u;ij}$ can be calculated by:
\begin{equation}
\begin{split}
    \boldsymbol{\Sigma}_{\boldsymbol{\eta}_{u;ij}} &=  \mathrm{E} \left\{ \boldsymbol{\eta}_{u;ij} \boldsymbol{\eta}_{u;ij}^T  \right\} \\
     &=  
    \boldsymbol{C}_{u;ij}^{\boldsymbol{v}_i} \boldsymbol{\Sigma}_{\boldsymbol{v}_i} {\boldsymbol{C}_{u;ij}^{\boldsymbol{v}_i T}} + 
    \boldsymbol{C}_{u;ij}^{\boldsymbol{w}_{ui}} \boldsymbol{\Sigma}_{\boldsymbol{w}_{ui}} {\boldsymbol{C}_{u;ij}^{\boldsymbol{w}_{ui} T}} \\
    & \quad \ \boldsymbol{C}_{u;ij}^{\boldsymbol{v}_j} \boldsymbol{\Sigma}_{\boldsymbol{v}_j} {\boldsymbol{C}_{u;ij}^{\boldsymbol{v}_j T}} + 
    \boldsymbol{C}_{u;ij}^{\boldsymbol{w}_{uj}} \boldsymbol{\Sigma}_{\boldsymbol{w}_{uj}} {\boldsymbol{C}_{u;ij}^{\boldsymbol{w}_{uj} T}} 
\end{split}
\label{eq68}
\end{equation}
\section{CRLBs for localization}  \label{appendix2}
Regarding the CRLBs for localization, we also consider the scenario with concurrent measurements. The CRLB for localization, as demonstrated in \cite{chan1994simple}, is derived for a vector of TDOA measurements with a covariance matrix $\boldsymbol{Q}$, where anchor 1 is selected as the reference anchor and all the TDOA estimates are defined between target $u$ and anchor pair $(1,i)$ ($i \in \mathcal{N}_a,i\neq 1$). We will not repeat the complete derivation here. However, it is important to note that the CRLB for localization is influenced by the covariance matrix of the TDOA estimates. Whether or not the polynomial TDOA model is applied during TDOA estimation will impact the covariance matrix of the TDOA estimates, thereby affecting the CRLB for localization.

For $\text{CRLB1}(\boldsymbol{p}_u)$, the polynomial TDOA model is not applied, so the CRLB for localization across different frames remains identical as the covariance matrix of the TDOA estimates, denoted as $\boldsymbol{Q}_u$, remains unchanged.
\begin{equation}
\boldsymbol{Q}_{u} = \frac{1}{2} 
 \left[  
\begin{matrix}
       2\sigma_n^2  & \sigma_n^2 & \cdots & \sigma_n^2\\
        \sigma_n^2  & 2\sigma_n^2 & \cdots & \sigma_n^2\\
         \vdots  & \vdots & \ddots & \sigma_n^2\\
          \sigma_n^2  & \sigma_n^2 & \cdots & 2\sigma_n^2
\end{matrix}
\right]
\label{eq69}
\end{equation}
where $\sigma_n^2$ is the variance of an independent TDOA measurement defined behind \eqref{eq39}.

However, for $\text{CRLB2}(\boldsymbol{p}_u)$, since the polynomial TDOA model is applied, the covariance matrix of the TDOA estimates may vary across different frames. We collect the diagonal entries of $\text{CRLB2}(\boldsymbol{\tau}_{u;ij})$ in \eqref{eq46} as
\begin{equation}
\begin{split}
    \left[{\sigma_n^{(1)}}^{2}, {\sigma_n^{(2)}}^{2}, \ldots,{\sigma_n^{(N_f)}}^{2} \right] = &\text{diag} 
\left (   \sigma_n^2 \boldsymbol{V}_u {\left( \boldsymbol{V}_u^T  \boldsymbol{V}_u \right)}^{-1}
     \boldsymbol{V}_u^T  \right )
\end{split}
\label{eq70}
\end{equation}
where ${\sigma_n^{(m)}}^{2}$ represents the variance of the TDOA estimate between target $u$ and an anchor pair in frame $m$ ($m=1,2,\ldots,N_f$). In this way, the covariance matrix of the vector of TDOA estimates in frame $m$ can be expressed as
\begin{equation}
\boldsymbol{Q}_{u}^{(m)} = \frac{1}{2} 
\left[\begin{matrix}
2{\sigma_n^{(m)}}^{2}  & {\sigma_n^{(m)}}^{2} & \cdots & {\sigma_n^{(m)}}^{2}\\
      {\sigma_n^{(m)}}^{2}  & 2{\sigma_n^{(m)}}^{2} & \cdots & {\sigma_n^{(m)}}^{2}\\
       \vdots  & \vdots & \ddots & {\sigma_n^{(m)}}^{2}\\
        {\sigma_n^{(m)}}^{2}  & {\sigma_n^{(m)}}^{2} & \cdots & 2{\sigma_n^{(m)}}^{2}
\end{matrix}\right]
\label{eq71}
\end{equation}
Using the newly derived covariance matrix of the TDOA measurements, we can determine the CRLB for localization in the corresponding frames.    

\end{appendices}

\bibliographystyle{IEEEtran} 
\bibliography{IEEEabrv,reference} 

\begin{IEEEbiography}[{\includegraphics[width=1in,height=1.25in,clip,keepaspectratio]{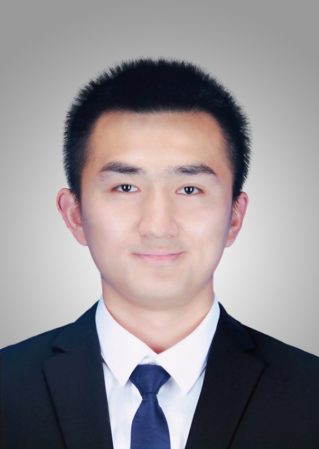}}]{Chenxin Tu} received the B.E. degree in electronic engineering from Tsinghua University, Beijing, China, in 2022, where he is currently pursuing the Ph.D. degree with the Department of Electronic Engineering.
His research interests include wireless localization and cooperative localization.
\end{IEEEbiography}

\begin{IEEEbiography}[{\includegraphics[width=1in,height=1.25in,clip,keepaspectratio]{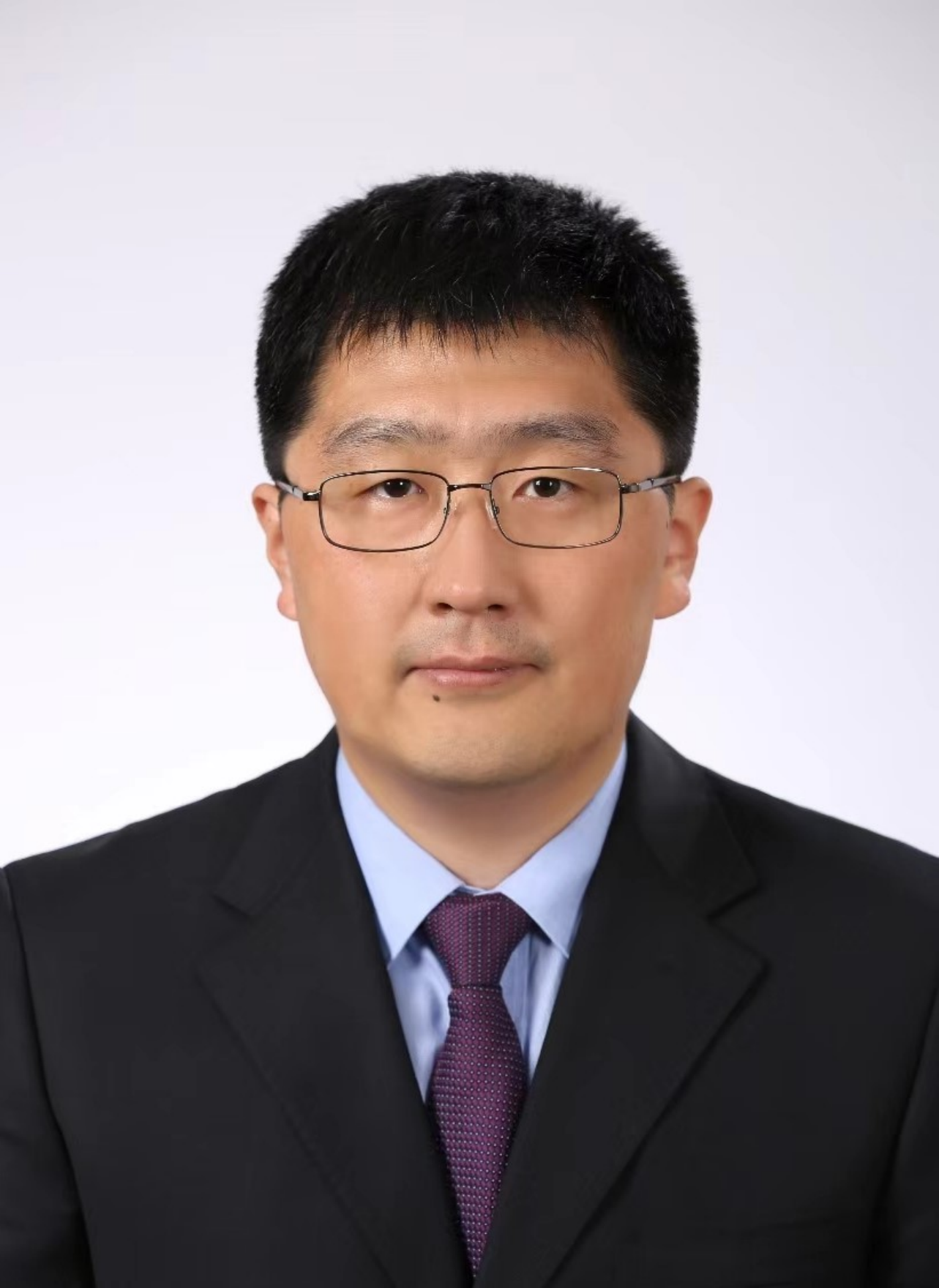}}]{Xiaowei Cui} received the B.S. and Ph.D. degrees in electronic engineering from Tsinghua University, Beijing, China, in 2000 and 2005, respectively. Since 2005, he has been with the Department of Electronic Engineering, Tsinghua University, where he is currently a Professor. 

His research interests include robust GNSS signal processing, multipath mitigation techniques, and high-precision positioning. He is a member of the Expert Group of China BeiDou Navigation Satellite System.
\end{IEEEbiography}

\begin{IEEEbiography}[{\includegraphics[width=1in,height=1.25in,clip,keepaspectratio]{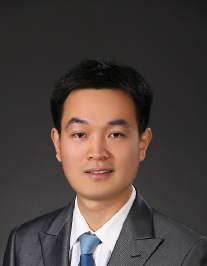}}]{Gang Liu} 
is an associate researcher at the Department of Electronic Engineering, Tsinghua University, China. His research interests include GNSS/INS integrated navigation techniques, and high-precision localization. He obtained both the BS and PhD degrees in instrument science and technology from Tsinghua University in 2007 and 2015, respectively.
\end{IEEEbiography}

\begin{IEEEbiography}[{\includegraphics[width=1in,height=1.25in,clip,keepaspectratio]{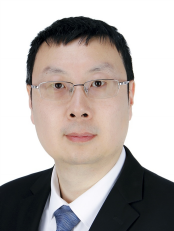}}]{Sihao Zhao}(Senior Member, IEEE)
received the B.S. and Ph.D. degrees in electronic engineering from Tsinghua University, Beijing, China, in 2005 and 2011, respectively. He is currently a Senior Staff Engineer with etherWhere Corporation, CA, USA. He was a Senior Algorithm Designer with NovAtel, Autonomy and Positioning division of Hexagon, Calgary, AB, Canada, from 2022 to 2024. From 2011 to 2013, he was an Electronics Systems Engineer with China Academy of Space Technology, Beijing, China. From 2013 to 2019, he was a Postdoctoral Researcher and then an Assistant Professor with the Department of Electronic Engineering, Tsinghua University, Beijing, China. From 2020 to 2021, he was a Research Associate with the Department of Electrical, Computer and Biomedical Engineering, Toronto Metropolitan University, Toronto, ON, Canada. He is an Editor of GPS Solutions.
\end{IEEEbiography}

\begin{IEEEbiography}[{\includegraphics[width=1in,height=1.25in,clip,keepaspectratio]{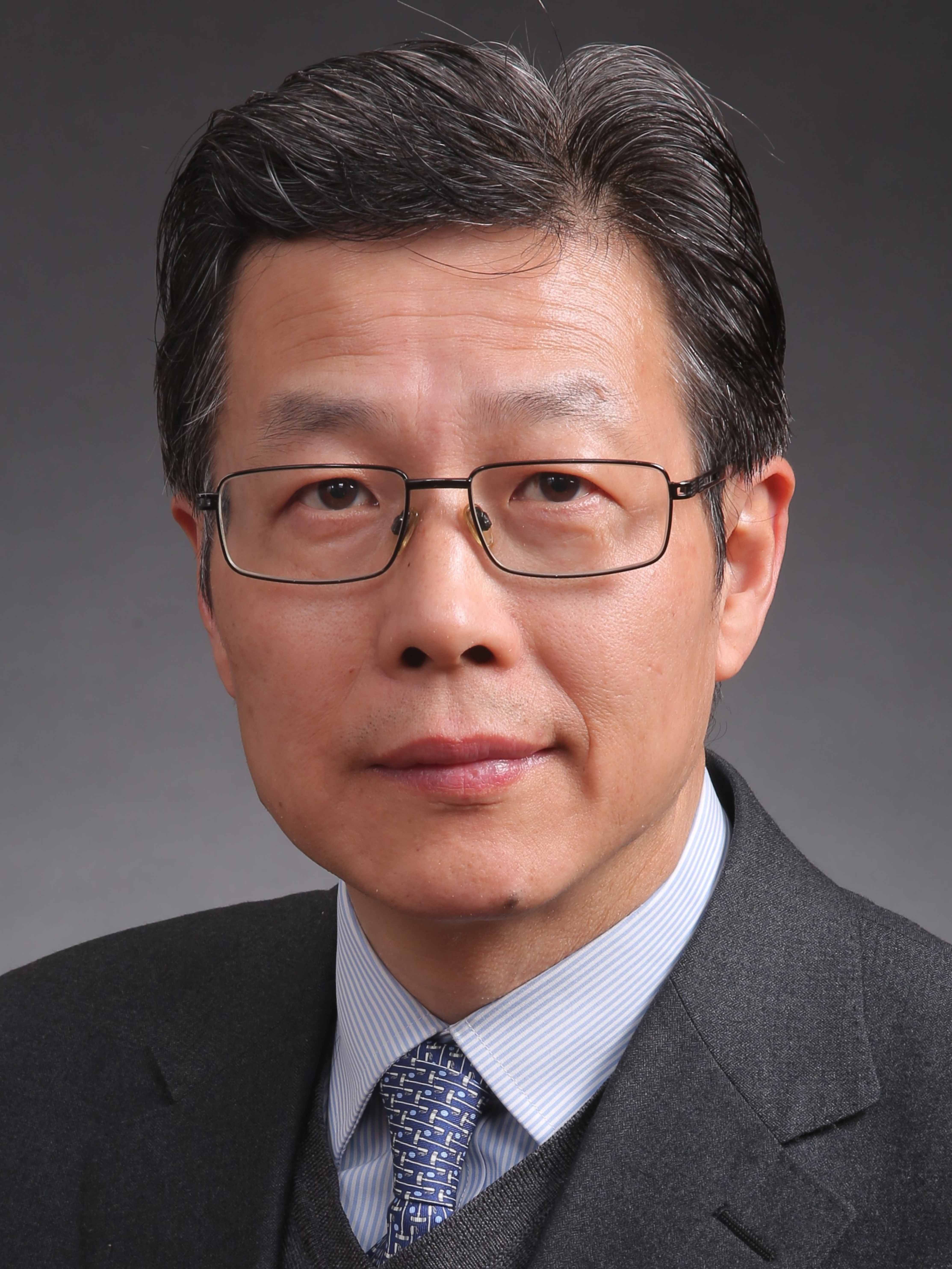}}]{Mingquan Lu} 
is a Professor in the Department of Electronic Engineering at Tsinghua University. He directs the PNT Research Center, which develops GNSS and alternative PNT technologies. His current research focuses on GNSS signal processing, GNSS receiver development and emerging PNT technologies. He authored or co-authored 5 books and book chapters, published over 300 journal and conference papers, and hold nearly 100 patents. He provided numerous services to the GNSS community, including associate editor for Satellite Navigation and Journal of Navigation and Positioning.

Dr. Lu is a fellow of ION and a recipient of the ION Thurlow Award. He provided numerous services to the GNSS community, including as an Associate Editor for Satellite Navigation and the
Journal of Navigation and Positioning.
\end{IEEEbiography}

\vfill

\end{document}